\documentclass[nofootinbib]{aastex631}
\usepackage{amsmath}	
\usepackage{booktabs}
\usepackage{float}
\usepackage{longtable} 

\begin{document}

\title{A sample study on the Fermi-LAT spectral variation behavior of blazars}

\correspondingauthor{Yunguo Jiang}
\email{jiangyg@sdu.edu.cn}
\author[0009-0007-1984-6603]{Junhao Deng}
\affiliation{Shandong Provincial Key Laboratory of Optical Astronomy and Solar-Terrestrial Environment, Institute of Space Sciences, Shandong University, Weihai, 264209, China}
\affiliation{School of Space Science and Technology, Shandong University, Weihai, 264209, China}
\author[0000-0003-2679-0445]{Yunguo Jiang}
\affiliation{Shandong Provincial Key Laboratory of Optical Astronomy and Solar-Terrestrial Environment, Institute of Space Sciences, Shandong University, Weihai, 264209, China}
\affiliation{School of Space Science and Technology, Shandong University, Weihai, 264209, China}



\begin{abstract}
In this study, we analyze the $\sim 16$ yr Fermi-LAT data of 160 blazars consisting of 92 flat-spectrum radio quasars (FSRQs), 64 BL Lacertae type objects (BL Lacs), and 4 blazar candidates of unknown type objects (BCU), and exhibit their spectral variation. By fitting a linear model ($y = kx + b$) to the monthly binned flux-index plots, we find that most of the FSRQs ($\sim 89\%$) display a harder-when-brighter (HWB) trend ($k>0$), while the majority of BL Lacs ($\sim 64\%$) exhibit a softer-when-brighter (SWB) trend ($k<0$). By plotting their slope ($k$) versus their variability index and inverse Compton (IC) peak frequency, we find a moderate positive correlation between $k$ and the variability index, and we find a moderate negative correlation between $k$ and the IC peak frequency, which we name as the gamma-ray variability sequence of blazars. Additionally, 7 blazars may not follow the linear variation trend, particularly in PKS 0537-441, AP Librae, and PKS 1510-089. Additional very-high-energy gamma-ray emission components may be the cause of the break in the variation trend.

\end{abstract}



\section{Introduction} \label{sec:intro}
Blazars are a subclass of active galactic nuclei (AGNs) with relativistic jets directed towards Earth \citep{1995PASP..107..803U}. Blazars are also can be divided into two subclasses, flat spectrum radio quasars (FSRQs) and BL Lacertae type objects (BL Lacs). The difference between them is that FSRQs exhibit pronounced broad emission lines in the optical spectrum and BL Lacs have no or very faint broad emission lines. This can be ascribed to the accretion disk in BL Lacs is not strong enough to activate the broad line region (BLR) \citep{2011MNRAS.414.2674G}.
Blazars are known to exhibit intrinsic variability at different timescales in multiple bands. Analyzing the spectral variation can help us understand the blazar's radiation process and components. In the optical range, \cite{2023MNRAS.519.5263Z} analyzed the spectral variation of 27 blazars and reported that almost all BL Lacs display the bluer-when-brighter (BWB) trend, while FSRQs often show the redder-when-brighter (RWB) behavior. This difference is attributed to the accretion disk in FSRQs, which contributes to the "big blue bump" in the optical UV range. Additionally, \cite{2023MNRAS.526.4040M} analyzed the long-term hard X-ray (14–195 keV) variability properties of 127 blazars and found that only five blazars exhibit significant spectral variation. They concluded that the variability in the hard X-ray is primarily driven by the processes affecting the particle injection. Currently, there are limited reports on the variation characteristics of blazars in the gamma-ray range, with only a few studies analyzing the spectral variation of individual targets. For instance, the gamma-ray spectral variation of ON 231, AO 0235-164, 1ES 1218+304, and PKS B1222+216 has been observed to follow the softer-when-brighter (SWB) trend \citep{2019ApJ...884...15S,2020ApJ...902...41W,2021MNRAS.508.1986C,2023MNRAS.524.4333D,2024ApJ...966...65W}. In contrast, some blazars (e.g., BL Lacertae, Ton 599, etc) have shown harder-when-brighter (HWB) behaviors \citep{2024ApJS..270...22W}. To understand the physical mechanisms of gamma-ray spectral variations, it is necessary to study a sample of blazars and investigate the variable characteristics in a systematic manner. Thus, in this paper, we analyze the Fermi-LAT spectral variation of 160 blazars and quantify their spectral behavior based on the variation trend. We aim to interpret these phenomena by proposing a unified scenario. This paper is organized as follows: Section \ref{sec:2} describes the data collection. Section \ref{sec:3} presents the analysis and results, and the summary is provided in Section \ref{sec:4}.

\section{Sample Selection and Data colloction} \label{sec:2}

The Fermi-LAT lightcurve repository\footnote{\url{https://fermi.gsfc.nasa.gov/ssc/data/access/lat/LightCurveRepository/}} (LCR) is a publicly available, continually updated library of gamma-ray lightcurves from Fermi-LAT sources \citep{2023ApJS..265...31A}. From the LCR, we collected the 30-day binned 0.1-100 GeV lightcurves and corresponding photon indices for 160 blazars with an average significance (AS) exceeding 50, using data spanning from August 5, 2008, to November 1, 2024. The sample comprises 92 FSRQs, 64 BL Lacs, and 4 blazar candidates of unknown type objects (BCUs). The AS for the blazars are provided directly by the LCR. We also collected the gamma-ray variability index ($V_{\rm{index}}$) of these blazars from the LCR. This index measures the degree of flux variation in the source, where a larger value indicates more significant variability. Additionally, we collected the inverse Compton (IC) peak frequency ($\nu^{\rm IC}_{p}$) of these blazars from \cite{2023ApJS..268...23F}.


\section{Analysis and Results} \label{sec:3}

\begin{figure}[h!]
	\centering
         \includegraphics[width=0.4\linewidth]{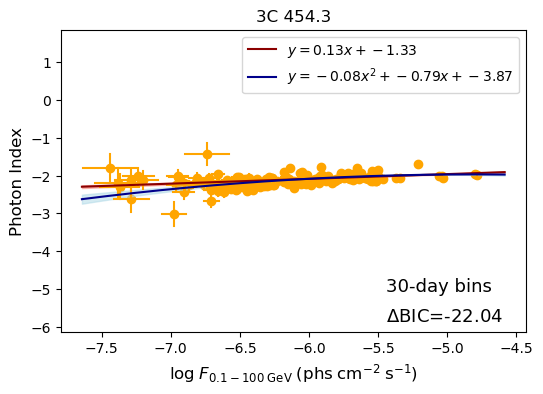}
    \caption{The photon index plotted against the 0.1-100 GeV flux for the brightest blazar 3C 454.3. The orange points represent the data with TS values greater than 10. The red line represents the fit of the linear function and the blue line represents the fit of the quadratic function. $\Delta \rm BIC$ is the compared fit result of the Bayesian Information Criterion between the two functions.} 
    \label{fig:indexplot}
\end{figure}

\begin{figure}[h!]
	\centering
         \includegraphics[width=0.4\linewidth]{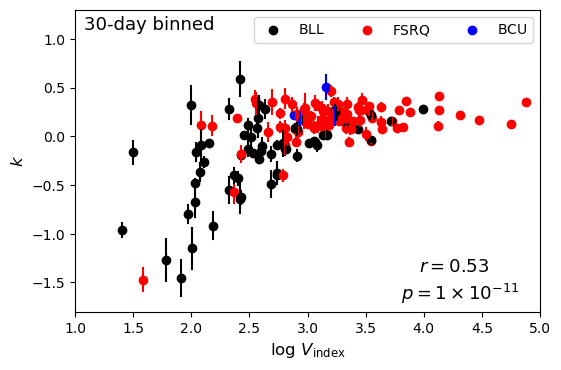}
         \includegraphics[width=0.4\linewidth]{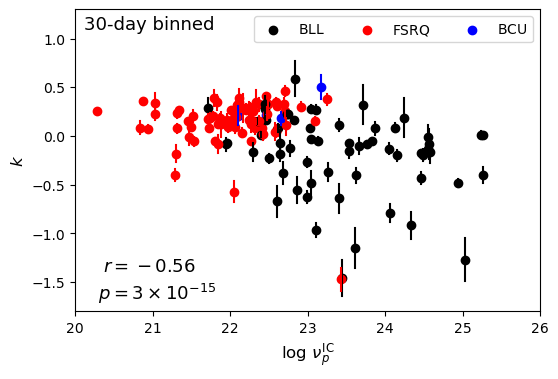}
         \includegraphics[width=0.4\linewidth]{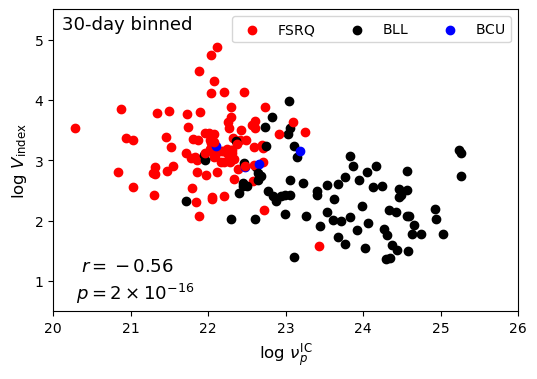}
    \caption{Top left panel: the variation slope ($k$) plotted against the gamma-ray variability index ($V_{\rm index}$). Top right panel: the variation slope ($k$) plotted against the IC peak frequency ($\nu^{\rm IC}_{p}$). Bottom panel: the gamma-ray variability index ($V_{\rm index}$) plotted against the IC peak frequency ($\nu^{\rm IC}_{p}$). All panels include the 160 blazars.} 
    \label{fig:sequence100}
\end{figure}

Figure \ref{fig:indexplot} displays the photon index plotted against the 0.1-100 GeV flux for the brightest blazar, 3C 454.3. For all blazars of our sample, the index-flux plots can be found in Appendix A. We fitted all the plots via the linear function ($y=kx+b$) with the Orthogonal Distance Regression (ODR) method, which can be done using \texttt{scipy.odr} module in Python. Table \ref{tab1} summarizes the slope ($k$) obtained from the best fit by the linear function. A positive value of $k$ indicates the HWB behavior, while a negative value indicates the SWB behavior. Among the 160 blazars, we found that the majority of BL Lacs (64\%) exhibit the SWB behavior, while almost all of the FSRQs (89\%) display HWB behavior. On the other hand, we plotted the distribution of $k$ against log $V_{\rm index}$ and log $\nu^{\rm IC}_{p}$ respectively, as shown in the top panels of Figure \ref{fig:sequence100}. For their correlation, we conducted statistics with Spearman’s coefficients, which can be done by using \texttt{scipy.stats} module in Python. In $k$ versus log $V_{\rm index}$ case, we found $r=0.53$ and $p=1\times10^{-11}$, indicating a moderate positive correlation. In $k$ versus log $\nu^{\rm IC}_{p}$ case, we found $r=-0.56$ and $p=3\times10^{-15}$, indicating a moderate negative correlation. In addition, the bottom panel of Figure \ref{fig:sequence100} also presents $V_{\rm index}$ plotted against $\nu^{\rm IC}_{p}$.

\begin{longtable}{cccccccc}
\caption{Monthly binned spectral variation trend of 160 blazars.} \label{tab1} \\
\hline
Object name & Class & AS & $V_{\rm index}$ & log $\nu^{IC}_{p}$ & Slope (30-d) & Slope error (30-d) & $\Delta \rm BIC$ \\
\hline
\endfirsthead
\hline
\endhead
\hline
\endfoot
TXS 0059+581 & FSRQ & 60.85 & 29.92 & 23.48 & 0.04 & 0.05 & -0.12 \\
4C +01.02 & FSRQ & 167.52 & 206.09 & 21.24 & 0.05 & 0.02 & -19.30 \\
S2 0109+22 & BLL & 126.49 & 151.23 & 23.67 & 0.08 & 0.05 & -5.96 \\
4C +31.03 & FSRQ & 67.61 & 1486.40 & 23.50 & 0.28 & 0.02 & -4.83 \\
PKS 0116-219 & FSRQ & 58.24 & 3494.66 & 22.99 & 0.03 & 0.08 & -7.26 \\
PKS 0130-17 & FSRQ & 60.93 & 32.62 & 25.42 & 0.41 & 0.09 & 1.03 \\
PKS 0131-522 & FSRQ & 54.94 & 42.54 & 23.99 & 0.48 & 0.06 & -5.15 \\
OC 457 & FSRQ & 72.21 & 50.79 & 22.72 & -0.05 & 0.07 & -4.42 \\
PKS 0139-09 & BLL & 58.08 & 267.60 & 22.18 & 0.20 & 0.08 & -5.68 \\
TXS 0141+268 & BLL & 100.82 & 53.11 & 22.23 & -0.57 & 0.14 & -759.96 \\
PKS 0208-512 & FSRQ & 120.41 & 22.18 & 20.90 & 0.16 & 0.02 & -3.18 \\
MG1 J021114+1051 & BLL & 85.53 & 33.33 & 21.79 & 0.11 & 0.06 & -58.11 \\
PKS 0215+015 & FSRQ & 63.78 & 23.17 & 23.76 & 0.04 & 0.09 & -2.55 \\
B2 0218+357 & FSRQ & 120.71 & 25.73 & 23.02 & -0.05 & 0.02 & -5.90 \\
3C 66A & BLL & 176.65 & 43.06 & 22.26 & -0.06 & 0.03 & 10.15 \\
PKS 0226-559 & FSRQ & 66.44 & 84.17 & 24.76 & 0.46 & 0.05 & -10.52 \\
4C +28.07 & FSRQ & 154.82 & 43.78 & 23.15 & 0.15 & 0.03 & -4.45 \\
PKS 0235+164 & BLL & 135.26 & 30.83 & 20.48 & 0.16 & 0.03 & -2.97 \\
PKS 0244-470 & FSRQ & 77.36 & 24.91 & 23.51 & 0.35 & 0.07 & 0.12 \\
PKS 0250-225 & FSRQ & 97.47 & 481.68 & 24.04 & 0.32 & 0.05 & -5.43 \\
PKS 0301-243 & BLL & 103.38 & 22.21 & 22.83 & 0.11 & 0.06 & -5.70 \\
PKS 0308-611 & FSRQ & 63.53 & 35.46 & 23.18 & 0.08 & 0.05 & -5.47 \\
PMN J0334-3725 & BLL & 71.60 & 34.73 & 22.04 & 0.32 & 0.20 & -118.93 \\
PKS 0332-403 & BLL & 82.81 & 159.13 & 23.22 & -0.18 & 0.07 & -12.02 \\
PKS 0336-01 & FSRQ & 99.61 & 134.29 & 24.71 & 0.26 & 0.07 & 4.69 \\
PKS 0402-362 & FSRQ & 137.66 & 626.21 & 22.58 & 0.27 & 0.02 & -2.27 \\
PKS 0420-01 & FSRQ & 65.96 & 4410.28 & 24.28 & 0.12 & 0.08 & -1.91 \\
PKS 0426-380 & BLL & 249.91 & 72.97 & 23.23 & -0.03 & 0.02 & 3.17 \\
PKS 0440-00 & FSRQ & 75.34 & 28.86 & 23.17 & 0.13 & 0.03 & 0.39 \\
PKS 0447-439 & BLL & 142.29 & 211.08 & 24.51 & -0.18 & 0.03 & -8.50 \\
PKS 0454-234 & FSRQ & 230.26 & 463.98 & 21.84 & 0.07 & 0.04 & -8.03 \\
S3 0458-02 & FSRQ & 90.91 & 40.63 & 21.93 & 0.04 & 0.05 & 3.57 \\
PKS 0502+049 & FSRQ & 83.29 & 422.54 & 24.91 & 0.27 & 0.03 & -20.35 \\
TXS 0506+056 & BLL & 105.02 & 56.88 & 21.24 & -0.05 & 0.04 & -11.92 \\
PKS 0507+17 & FSRQ & 62.85 & 153.75 & 21.48 & 0.33 & 0.06 & -3.83 \\
PKS 0516-621 & BLL & 57.08 & 23.12 & 23.05 & -0.55 & 0.13 & 0.69 \\
TXS 0518+211 & BLL & 119.48 & 102.02 & 22.69 & -0.20 & 0.06 & -65.97 \\
PKS 0524-485 & FSRQ & 60.94 & 47.16 & 21.55 & 0.33 & 0.07 & -4.80 \\
OG 050 & FSRQ & 76.83 & 22.99 & 20.83 & 0.09 & 0.05 & 19.02 \\
TXS 0529+483 & FSRQ & 54.02 & 82.62 & 23.34 & -0.57 & 0.11 & -50.74 \\
PKS 0537-441 & BLL & 221.65 & 38.57 & 23.96 & 0.28 & 0.04 & 43.58 \\
TXS 0603+476 & BLL & 61.74 & 33.19 & 22.19 & -1.15 & 0.21 & -91.11 \\
B3 0609+413 & BLL & 78.58 & 21.85 & 21.60 & -1.45 & 0.19 & -82.31 \\
PKS 0700-661 & BLL & 90.26 & 2733.37 & 22.86 & -0.64 & 0.15 & -749.16 \\
B2 0716+33 & FSRQ & 76.08 & 134.42 & 22.44 & 0.26 & 0.07 & -8.48 \\
S5 0716+71 & BLL & 298.17 & 142.33 & 23.22 & 0.08 & 0.03 & -3.22 \\
4C +14.23 & FSRQ & 77.22 & 34.70 & 22.33 & 0.09 & 0.06 & -9.93 \\
PKS 0727-11 & FSRQ & 132.11 & 31.18 & 20.67 & -0.05 & 0.03 & 6.09 \\
PKS 0735+17 & BLL & 77.21 & 34.82 & 22.22 & -0.07 & 0.03 & -4.15 \\
PKS 0736+01 & FSRQ & 87.06 & 21.92 & 22.84 & 0.25 & 0.04 & -3.49 \\
GB6 J0742+5444 & FSRQ & 66.11 & 41.57 & 24.27 & 0.09 & 0.06 & -0.11 \\
PKS 0805-07 & FSRQ & 94.42 & 36.25 & 23.87 & 0.35 & 0.04 & -10.15 \\
1ES 0806+524 & BLL & 93.65 & 23.80 & 22.27 & -0.01 & 0.12 & -178.21 \\
OJ 014 & BLL & 70.06 & 26.87 & 22.02 & -0.49 & 0.13 & -910.15 \\
S4 0814+42 & BLL & 105.28 & 427.68 & 21.87 & -0.62 & 0.07 & -7.62 \\
PKS 0823-223 & BLL & 79.40 & 71.14 & 23.10 & -0.40 & 0.08 & -3.02 \\
PKS 0829+046 & BLL & 70.22 & 24.39 & 19.64 & 0.28 & 0.09 & -3.08 \\
4C +71.07 & FSRQ & 75.82 & 644.25 & 23.22 & 0.25 & 0.03 & -3.04 \\
GB6 J0850+4855 & BLL & 55.95 & 24.46 & 22.12 & 0.59 & 0.18 & -133.59 \\
PMN J0850-1213 & FSRQ & 50.94 & 40.90 & 23.60 & 0.39 & 0.10 & 4.18 \\
OJ 287 & BLL & 100.24 & 76.31 & 21.74 & 0.08 & 0.04 & 9.79 \\
PKS 0903-57 & BCU & 68.35 & 39.03 & 22.71 & 0.22 & 0.02 & 17.44 \\
PKS 0907-023 & FSRQ & 51.82 & 23.99 & 22.83 & 0.11 & 0.10 & -3.15 \\
S4 0917+44 & FSRQ & 66.23 & 203.85 & 25.00 & 0.13 & 0.04 & -15.53 \\
OK 630 & FSRQ & 79.72 & 91.15 & 21.61 & 0.33 & 0.06 & -6.03 \\
4C +55.17 & FSRQ & 180.73 & 243.13 & 23.37 & -1.47 & 0.12 & -20.49 \\
S4 0954+65 & BLL & 87.45 & 1322.41 & 21.68 & 0.15 & 0.03 & -16.78 \\
PKS 1004-217 & FSRQ & 50.44 & 84.11 & 24.13 & 0.29 & 0.06 & 4.41 \\
1H 1013+498 & BLL & 158.52 & 490.36 & 21.42 & -0.19 & 0.06 & -7.38 \\
S4 1030+61 & FSRQ & 87.89 & 833.20 & 21.60 & 0.04 & 0.05 & -0.52 \\
GB6 J1037+5711 & BLL & 91.86 & 153.19 & 21.74 & -0.92 & 0.14 & -7.44 \\
S5 1044+71 & FSRQ & 155.43 & 64.50 & 24.13 & 0.25 & 0.03 & -5.27 \\
4C +01.28 & BLL & 94.76 & 23.48 & 23.58 & -0.23 & 0.05 & -3.40 \\
PKS B1056-113 & BLL & 68.44 & 174.73 & 23.42 & -0.37 & 0.09 & -258.74 \\
PKS 1101-536 & BLL & 56.20 & 505.80 & 22.64 & -0.38 & 0.11 & -162.12 \\
Mkn 421 & BLL & 325.52 & 37.27 & 20.12 & 0.01 & 0.03 & -6.43 \\
PKS 1124-186 & FSRQ & 112.72 & 43.50 & 21.03 & 0.38 & 0.06 & -6.85 \\
PKS B1130+008 & BLL & 53.66 & 22.56 & 24.50 & -0.96 & 0.07 & -4.17 \\
S4 1144+40 & FSRQ & 99.36 & 30.87 & 25.47 & 0.11 & 0.05 & -7.81 \\
PKS 1144-379 & BLL & 51.10 & 182.52 & 25.12 & -0.67 & 0.16 & -305.42 \\
Ton 599 & FSRQ & 183.71 & 40.73 & 23.11 & 0.28 & 0.03 & 14.15 \\
TXS 1206+549 & FSRQ & 53.53 & 115.64 & 21.12 & 0.34 & 0.11 & -4.49 \\
B2 1215+30 & BLL & 134.95 & 356.95 & 23.34 & -0.13 & 0.06 & -19.84 \\
S5 1217+71 & FSRQ & 65.98 & 24.84 & 22.25 & 0.09 & 0.06 & -3.99 \\
PG 1218+304 & BLL & 69.29 & 25.86 & 22.04 & -1.27 & 0.22 & -243.49 \\
W Comae & BLL & 76.83 & 219.92 & 25.10 & -0.13 & 0.08 & -12.13 \\
4C +21.35 & FSRQ & 215.40 & 60.06 & 21.63 & 0.22 & 0.03 & -11.46 \\
3C 273 & FSRQ & 130.92 & 178.75 & 20.56 & 0.36 & 0.03 & -2.92 \\
ON 246 & BLL & 83.29 & 26.21 & 22.27 & 0.27 & 0.05 & -7.21 \\
MG1 J123931+0443 & FSRQ & 104.27 & 7045.78 & 23.05 & 0.19 & 0.04 & -5.65 \\
PKS 1244-255 & FSRQ & 108.87 & 1731.69 & 20.04 & 0.21 & 0.04 & -4.24 \\
PG 1246+586 & BLL & 116.68 & 236.17 & 23.92 & -0.79 & 0.09 & -17.93 \\
3C 279 & FSRQ & 291.63 & 1080.20 & 24.18 & 0.17 & 0.02 & -4.21 \\
TXS 1318+225 & FSRQ & 52.34 & 128.33 & 24.02 & 0.39 & 0.08 & 0.07 \\
PMN J1329-5608 & BLL & 53.93 & 35.26 & 20.62 & -0.07 & 0.02 & -3.75 \\
PKS 1329-049 & FSRQ & 54.50 & 34.86 & 21.94 & 0.23 & 0.06 & -3.26 \\
PMN J1332-1256 & FSRQ & 53.14 & 351.22 & 20.83 & -0.01 & 0.08 & -8.66 \\
B3 1343+451 & FSRQ & 155.21 & 38.20 & 22.16 & 0.30 & 0.04 & -9.34 \\
87GB 141615.9+355650 & BCU & 65.39 & 34.66 & 23.14 & 0.50 & 0.12 & -0.97 \\
NVSS J141922-083830 & FSRQ & 54.40 & 41.13 & 21.32 & 0.35 & 0.12 & 2.44 \\
PKS 1424+240 & BLL & 156.35 & 28.01 & 22.87 & -0.43 & 0.06 & -6.00 \\
PKS 1424-41 & FSRQ & 293.39 & 39.85 & 22.83 & 0.41 & 0.03 & -4.02 \\
PKS 1440-389 & BLL & 71.16 & 175.75 & 24.10 & -0.16 & 0.12 & -204.11 \\
PKS 1441+25 & FSRQ & 53.85 & 42.27 & 24.03 & 0.38 & 0.05 & -4.35 \\
TXS 1452+516 & BLL & 80.80 & 57.00 & 22.83 & -0.15 & 0.08 & -19.86 \\
PKS 1454-354 & FSRQ & 64.84 & 24.87 & 21.97 & 0.16 & 0.06 & -3.23 \\
PKS 1502+106 & FSRQ & 230.28 & 28.30 & 22.05 & 0.27 & 0.03 & -1.93 \\
B2 1504+37 & FSRQ & 62.18 & 31.61 & 21.52 & 0.21 & 0.06 & -4.00 \\
PKS 1510-089 & FSRQ & 238.71 & 76.05 & 23.85 & 0.08 & 0.02 & 157.08 \\
AP Librae & BLL & 94.96 & 55.37 & 20.67 & -0.26 & 0.05 & 22.88 \\
B2 1520+31 & FSRQ & 180.73 & 37.65 & 21.33 & 0.19 & 0.05 & -37.41 \\
GB6 J1542+6129 & BLL & 135.32 & 40.07 & 23.59 & -0.10 & 0.09 & 3.77 \\
PG 1553+113 & BLL & 175.39 & 5930.06 & 21.60 & -0.48 & 0.04 & -1.64 \\
GB6 J1604+5714 & FSRQ & 52.69 & 147.72 & 23.25 & 0.12 & 0.13 & -169.99 \\
4C +38.41 & FSRQ & 165.54 & 559.06 & 24.79 & 0.10 & 0.02 & -8.47 \\
Mkn 501 & BLL & 161.70 & 30.85 & 20.70 & -0.40 & 0.08 & -664.38 \\
TXS 1700+685 & FSRQ & 79.41 & 34.65 & 21.80 & 0.28 & 0.05 & -13.28 \\
B3 1708+433 & FSRQ & 66.58 & 24.14 & 21.02 & 0.17 & 0.08 & -4.98 \\
PKS 1716-771 & BCU & 57.45 & 6652.58 & 19.46 & 0.18 & 0.07 & -2.25 \\
S4 1726+45 & FSRQ & 60.63 & 31.52 & 23.17 & 0.24 & 0.05 & -0.14 \\
B2 1732+38A & FSRQ & 65.68 & 32.27 & 22.69 & 0.26 & 0.06 & -4.28 \\
4C +51.37 & FSRQ & 50.61 & 148.05 & 21.47 & -0.05 & 0.05 & -3.61 \\
S4 1749+70 & BLL & 102.63 & 49.53 & 20.53 & -0.08 & 0.04 & -3.58 \\
OT 081 & BLL & 72.42 & 25.92 & 20.87 & 0.17 & 0.03 & 2.46 \\
RX J1754.1+3212 & BLL & 60.45 & 30.31 & 21.86 & 0.19 & 0.20 & -10.04 \\
S5 1803+784 & BLL & 120.66 & 27.17 & 23.05 & 0.01 & 0.02 & -16.75 \\
PMN J1802-3940 & FSRQ & 65.03 & 32.50 & 21.80 & 0.02 & 0.06 & 0.41 \\
3C 371 & BLL & 82.88 & 27.65 & 21.32 & -0.16 & 0.09 & -258.49 \\
4C +56.27 & BLL & 58.48 & 33.28 & 23.53 & 0.28 & 0.10 & -100.48 \\
PKS 1824-582 & FSRQ & 56.82 & 57.06 & 23.42 & 0.07 & 0.04 & -3.68 \\
PKS 1830-211 & FSRQ & 126.79 & 179.38 & 23.75 & 0.16 & 0.02 & 3.77 \\
S4 1849+67 & FSRQ & 80.00 & 797.25 & 23.88 & 0.16 & 0.04 & -3.59 \\
PKS B1908-201 & FSRQ & 63.41 & 541.29 & 22.51 & -0.40 & 0.06 & 1.93 \\
PKS 1936-623 & BLL & 72.72 & 41.55 & 23.73 & -0.09 & 0.06 & -4.55 \\
PKS 1954-388 & FSRQ & 57.60 & 396.31 & 23.78 & 0.19 & 0.03 & -4.95 \\
1ES 1959+650 & BLL & 151.03 & 23.20 & 23.82 & 0.02 & 0.04 & -7.21 \\
S5 2007+77 & BLL & 61.50 & 253.22 & 22.94 & 0.33 & 0.09 & -23.08 \\
PKS 2023-07 & FSRQ & 126.54 & 617.93 & 22.75 & 0.31 & 0.04 & 10.90 \\
PKS 2052-47 & FSRQ & 102.74 & 24.53 & 25.32 & 0.18 & 0.04 & -2.67 \\
B2 2113+29 & FSRQ & 54.25 & 1214.60 & 22.65 & -0.08 & 0.09 & -9.27 \\
MH 2136-428 & BLL & 96.85 & 22.36 & 22.52 & 0.08 & 0.07 & -6.37 \\
PMN J2141-6411 & BCU & 69.68 & 38.11 & 22.79 & 0.21 & 0.09 & -4.30 \\
OX 169 & FSRQ & 92.65 & 22.07 & 23.18 & -0.18 & 0.08 & -5.98 \\
PKS 2142-75 & FSRQ & 97.41 & 75.80 & 21.50 & 0.27 & 0.07 & -4.15 \\
PKS 2155-304 & BLL & 224.64 & 55.88 & 24.74 & -0.13 & 0.03 & -1.92 \\
BL Lac & BLL & 204.12 & 36.50 & 21.81 & 0.22 & 0.02 & -2.46 \\
PKS 2201+171 & FSRQ & 61.97 & 50.78 & 22.60 & 0.29 & 0.15 & 1.75 \\
PKS 2227-08 & FSRQ & 64.74 & 25.36 & 22.90 & 0.08 & 0.07 & -0.22 \\
CTA 102 & FSRQ & 349.94 & 79.11 & 22.61 & 0.35 & 0.01 & 8.82 \\
PKS 2233-148 & BLL & 100.77 & 33.43 & 20.47 & 0.22 & 0.04 & 1.09 \\
RGB J2243+203 & BLL & 79.00 & 49.36 & 24.56 & -0.09 & 0.15 & -38.99 \\
TXS 2241+406 & FSRQ & 97.12 & 235.66 & 21.81 & 0.15 & 0.04 & 4.25 \\
PMN J2250-2806 & BLL & 57.24 & 111.23 & 23.03 & -0.07 & 0.13 & -1.77 \\
3C 454.3 & FSRQ & 465.15 & 35.08 & 21.91 & 0.13 & 0.02 & -22.04 \\
PKS 2255-282 & FSRQ & 58.67 & 53.85 & 24.04 & 0.11 & 0.03 & -7.06 \\
B2 2308+34 & FSRQ & 85.81 & 293.96 & 23.31 & 0.20 & 0.05 & -4.88 \\
B2 2319+31 & FSRQ & 63.13 & 71.57 & 24.58 & 0.15 & 0.06 & -4.25 \\
PKS 2320-035 & FSRQ & 80.08 & 51.98 & 25.03 & 0.23 & 0.07 & -2.00 \\
PKS 2326-502 & FSRQ & 183.24 & 99.20 & 21.31 & 0.11 & 0.04 & -9.75 \\
PMN J2345-1555 & FSRQ & 132.25 & 30.45 & 24.66 & 0.30 & 0.04 & -2.90 \\
\end{longtable}

The spectral energy distribution (SED) of blazars is often well-reproduced by the logarithmic parabola function. \cite{2024ApJ...966...65W} indicated that the spectral variability slope ($k$) in the shock-in-jet model could be given by: 
\begin{equation}
k=[1/(2b)+{\rm log}(\nu/\delta \nu'_p)]^{-1},
\label{eqk}
\end{equation}
where $\nu$ is the observed frequency, $\delta$ is the Doppler factor, $\nu'_p$ is the intrinsic peak frequency, and $b$ is the curvature index. \cite{2014ApJ...788..179C} studied the multi-band SED of 48 blazars and presented the curvature index of their IC bump. In their sample, the curvature index ranged from 0.032 to 0.48. In addition, the main observed energy range of the Fermi-LAT is 0.1-300 GeV. Thus, we set $\nu$ to be $10^{23}$ Hz, which is the most typical frequency of Fermi-LAT. For the FSRQ objects, we fix $\delta \nu'_p = 10^{21}$ Hz. Substituting the values of $b$ and $\delta \nu'_p$ into Equation \ref{eqk}, we obtained that $k$ is from 0.05 to 0.34, which is consistent with the observation. Similarly, for the BL Lac objects, we set $\delta \nu'_p = 10^{25}$ Hz, Equation \ref{eqk} predicts a negative $k$ when $b>0.25$. However, BL Lacs generally exhibit a small curvature index, which rarely exceeds 0.25. This suggests that Equation \ref{eqk} does not sufficiently explain why the majority of BL Lacs display SWB behavior. The radiation from BL Lacs may result from the superposition of multiple non-thermal components within the jet, as inferred from the low polarization degrees of BL Lacs \citep{2016MNRAS.462.4267J}. Since Equation \ref{eqk} considers only a single-component scenario, it is inapplicable to BL Lacs.

\begin{figure}[h!]
	\centering
         \includegraphics[width=0.4\linewidth]{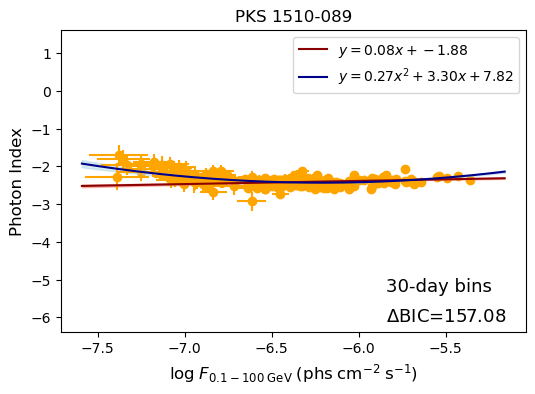}
         \includegraphics[width=0.4\linewidth]{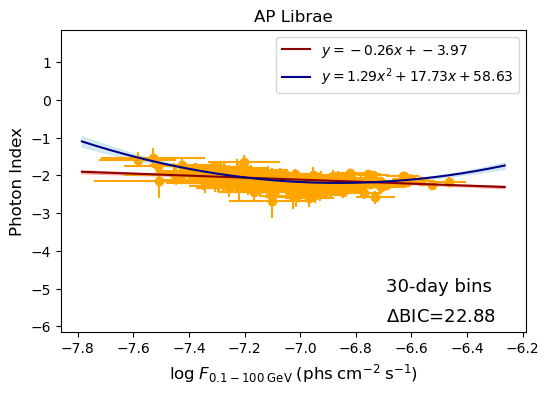}
         \includegraphics[width=0.4\linewidth]{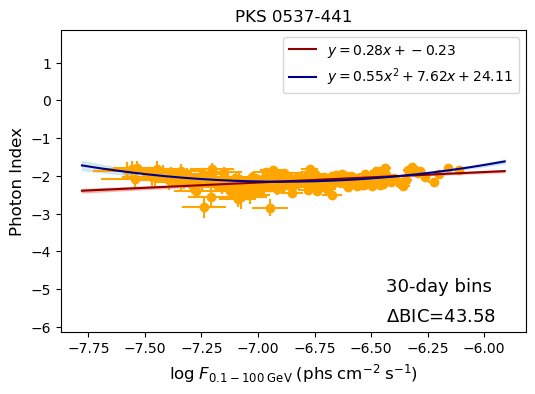}
    \caption{The photon index versus 0.1-100 GeV flux plot for PKS 1510-089, PKS 0537-441, and AP Librae, follows the same instructions as Figure \ref{fig:indexplot}.} 
    \label{fig:indexplot2}
\end{figure}

On the other hand, we noticed that the linear function provided a poor fit for some blazars because their variation behavior change as flux increases, such as changing from the SWB trend at the low flux period to the HWB trend at the high flux period. Thus, we also applied the quadratic function to all blazars of the sample and compared the fitting results of the linear and quadratic models using the Bayesian Information Criterion (BIC). By calculating the BIC values, we can quantify the relative goodness of fit and determine whether the quadratic function offers a statistically better representation of the data than the linear function. If the difference between the BIC values of the linear fit and quadratic fit is significantly less than 0 ($\Delta \text{BIC}<0$), the linear model provides a better fit. Conversely, when $\Delta \text{BIC}$ is significantly greater than 0, the quadratic model is favored. If $\Delta \text{BIC}$ is close to 0, both models provide comparable fits. In the 30-day binned scenario, among the 160 blazars, 7 blazars (3C 66A, OG 050, PKS 0537-441, PKS 0903-57, Ton 599, PKS 1510-089, AP Librae, and PKS 2023-07) have $\Delta \text{BIC} > 10$, where 3 blazars (PKS 1510-089, AP Librae, and PKS 0537-441) have $\Delta \text{BIC} > 20$, as shown in Figure \ref{fig:indexplot2}. PKS 1510-089 shows the most pronounced break, with the largest $\Delta \text{BIC}$ value of 149.36.

Some studies reported that in PKS 1510-089 and AP Librae, the high-energy (HE, $<100$ GeV) and very-high-energy (VHE, $>100$ GeV) gamma-ray emissions may not be produced co-spatially, which means that there are at least two gamma-ray emission components (a HE component and a VHE component). For instance, \cite{2023ApJ...952L..38A} reported that in July 2021, PKS 1510-089 exhibited a significant flux drop in the Fermi-LAT observation range, which persisted throughout 2022. However, in the H.E.S.S observation range, the VHE flux did not show significant changes. They suggested that the primary emission region at the base of the jet responsible for the HE component had vanished in 2022, while the emission zone far from the black hole, responsible for the VHE component, remained unchanged. On the other hand, \cite{2024MNRAS.532.1991B} observed that in 2022, PKS 1510-089 experienced a sudden drop in optical polarization and a significant shift in its frequency dependence, providing additional evidence for the two emission components. For AP Librae, \cite{2016A&A...588A.110Z} reported that the broadband SED of AP Librae could not be reproduced by the zone located close to the black hole. \cite{2022ApJ...924...57R} conducted a detailed study of the multi-wavelength imaging of AP Librae and concluded that the zone dominates the VHE emission of AP Librae at hundreds of parsecs from the black hole. In addition, AP Librae also exhibits unusual polarization behavior in the optical band. In Figure 3 of \cite{2024MNRAS.532.1991B}, it can be seen that when the flux of AP Librae in the optical and Fermi-LAT energy ranges increases, the degree of optical polarization decreases, and conversely, when the flux decreases, the degree of optical polarization increases. This unusual polarization behavior may indicate that AP Librae also has multiple non-thermal components within its jet. Thus, we speculate that the additional emission component is the cause of the unusual variation behavior. When the HE component was at a low flux level the VHE component dominated the Fermi-LAT band, and the SWB behavior could be observed if the HE component had a greater contribution toward the total Fermi-LAT range flux during brightening. On the other hand, when the HE component is at a high flux level and the contribution from the VHE component could be negligible, the spectral behavior will change to HWB similar to other normal blazars with the HWB trend. For PKS 0537-441, future studies could involve a detailed analysis of its broadband SED and lightcurve to determine whether an additional VHE component is also present.

\section{Summary} \label{sec:4}

In this work, we analyzed the Fermi-LAT spectral variation of 160 blazars, focusing on the behavior of their photon index as the function of flux. Our main findings are as follows:

1. In the monthly binned analysis, by the linear model fit, we found that almost all of the FSRQs (89\%) display the HWB trend, while the majority of BL Lacs (64\%) exhibit the SWB trend. The HWB trend in FSRQs could be explained by the shock-in-jet model, but the SWB trend in BL Lacs is still an open question.

2. By plotting the slope $k$ versus the variability index and IC peak frequency, we find a moderate positive correlation between $k$ and the variability index, and we find a moderate negative correlation between $k$ and the IC peak frequency, which we name as the gamma-ray variability sequence of blazars. 


3. In the sample, 7 blazars may not follow the linear variation trend. This was particularly evident in blazars like PKS 0537-441, AP Librae, and PKS 1510-089. The presence of multiple gamma-ray emission components may be the cause for the presence of the break in the variation trend.


\begin{acknowledgments}
This work has been funded by the National Natural Science Foundation of China under grant no. U2031102.
\end{acknowledgments}

%

\vspace{5mm}
\facilities{Fermi(LAT)}


\software{SciPy \citep{2020NatMe..17..261V}}


\appendix
\counterwithin{figure}{section}
\renewcommand{\thefigure}{A\arabic{figure}}
\section{Photon index vs. flux for all targets}\label{sec:appA}
\begin{figure}[h!]
	\centering
\includegraphics[width=0.267\linewidth]{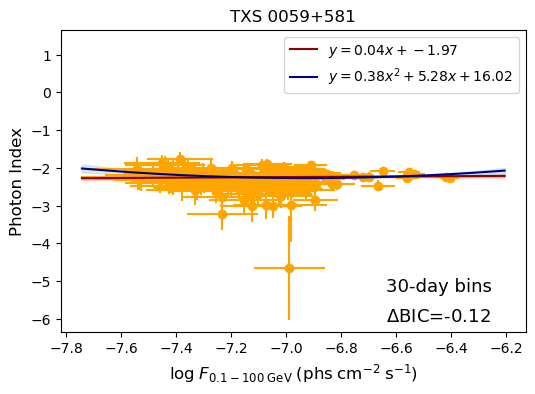}
\includegraphics[width=0.267\linewidth]{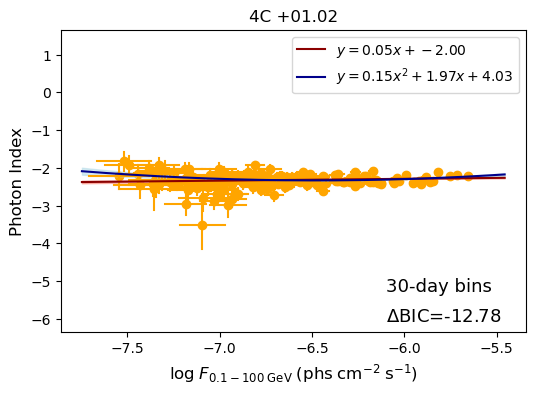}
\includegraphics[width=0.267\linewidth]{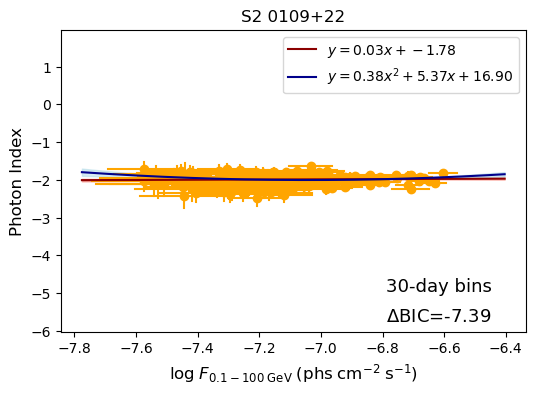}
\includegraphics[width=0.267\linewidth]{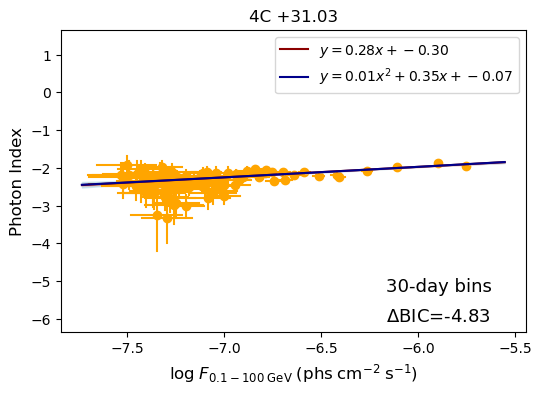}
\includegraphics[width=0.267\linewidth]{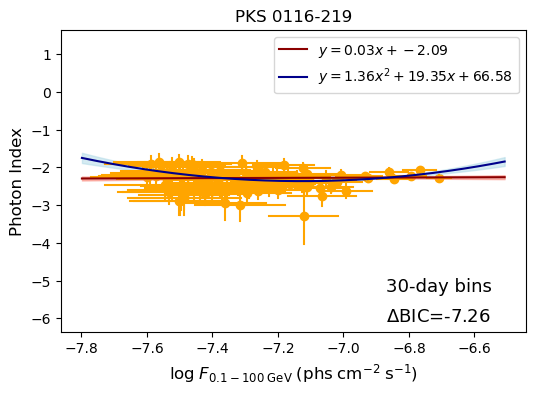}
\includegraphics[width=0.267\linewidth]{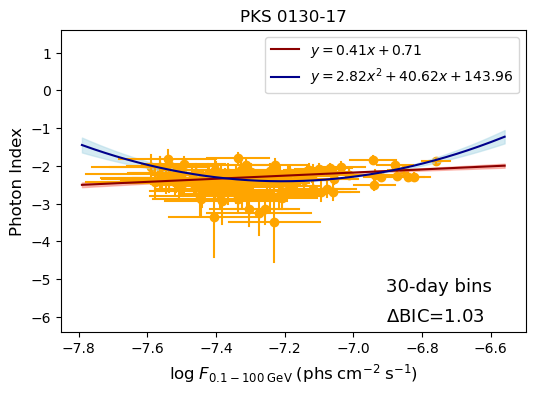}
\includegraphics[width=0.267\linewidth]{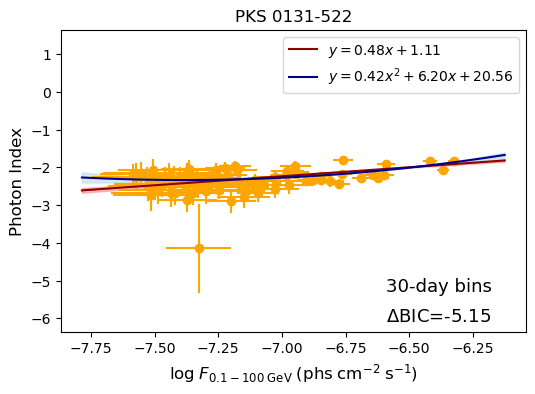}
\includegraphics[width=0.267\linewidth]{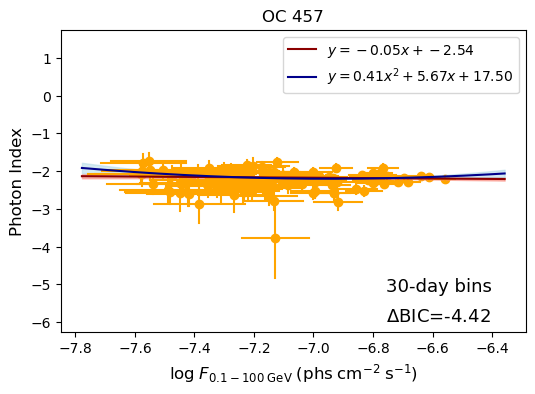}
\includegraphics[width=0.267\linewidth]{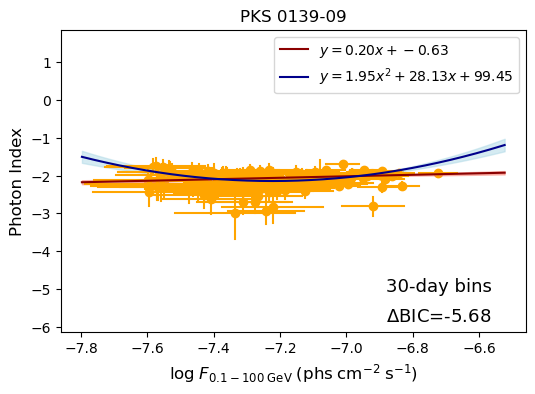}
\includegraphics[width=0.267\linewidth]{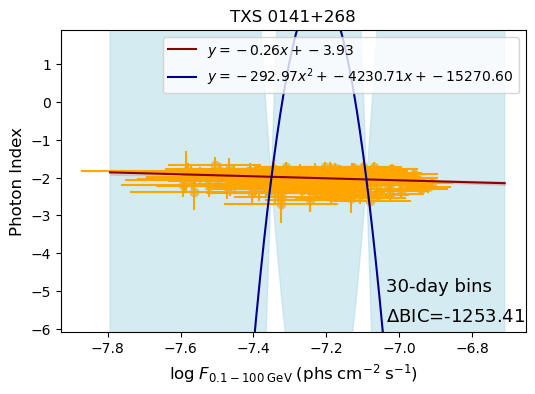}
\includegraphics[width=0.267\linewidth]{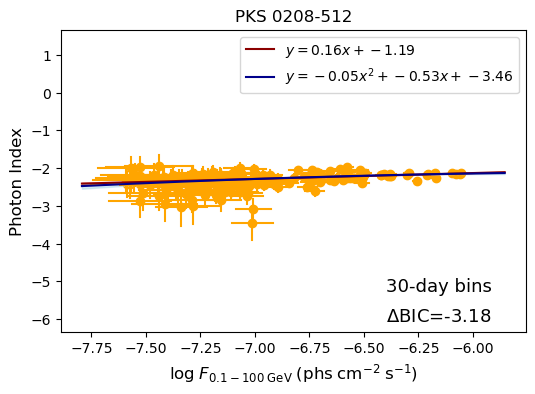}
\includegraphics[width=0.267\linewidth]{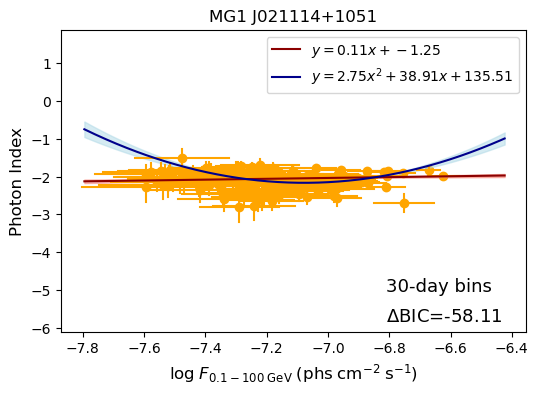}
\includegraphics[width=0.267\linewidth]{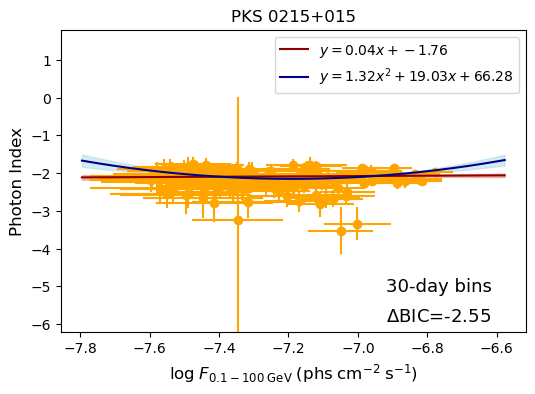}
\includegraphics[width=0.267\linewidth]{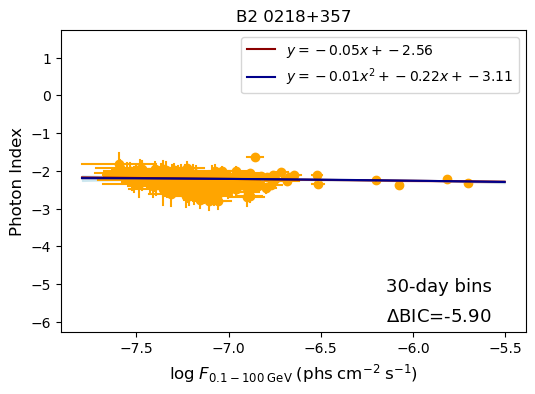}
\includegraphics[width=0.267\linewidth]{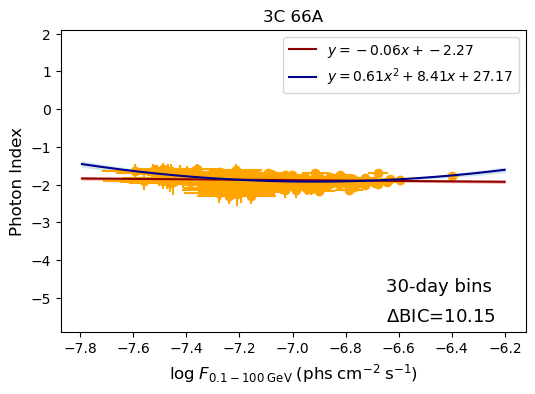}
\includegraphics[width=0.267\linewidth]{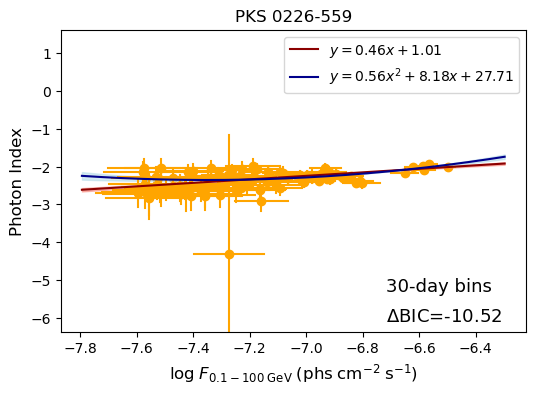}
\includegraphics[width=0.267\linewidth]{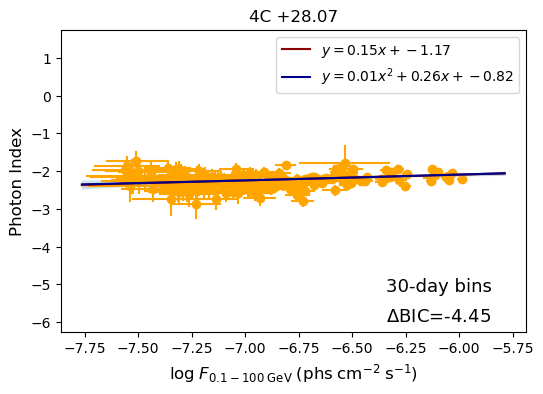}
\includegraphics[width=0.267\linewidth]{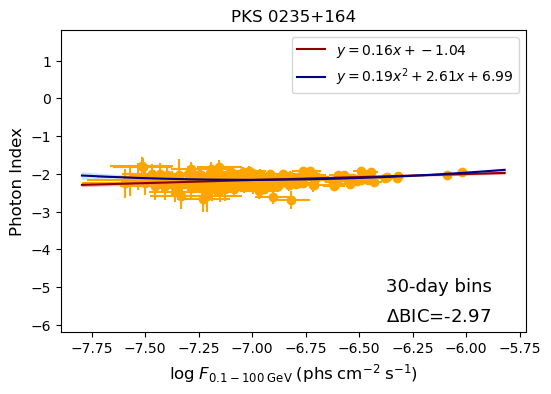}
    \caption{The same as in Figure \ref{fig:indexplot}.} 
    \label{fig:a1}
\end{figure}

\begin{figure}[h!]
	\centering
\includegraphics[width=0.267\linewidth]{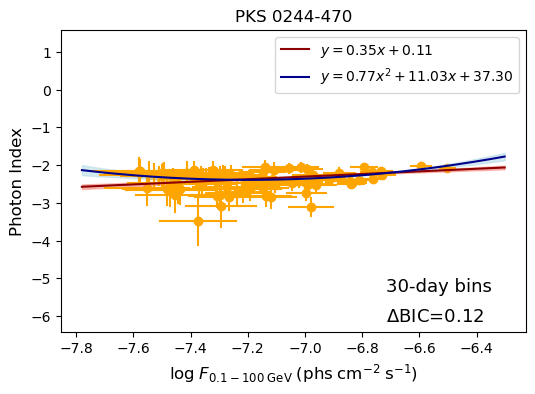}
\includegraphics[width=0.267\linewidth]{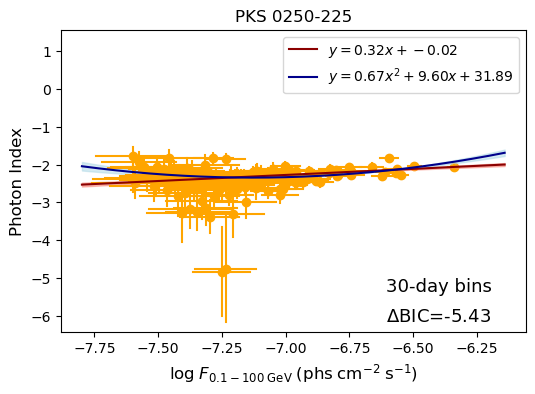}
\includegraphics[width=0.267\linewidth]{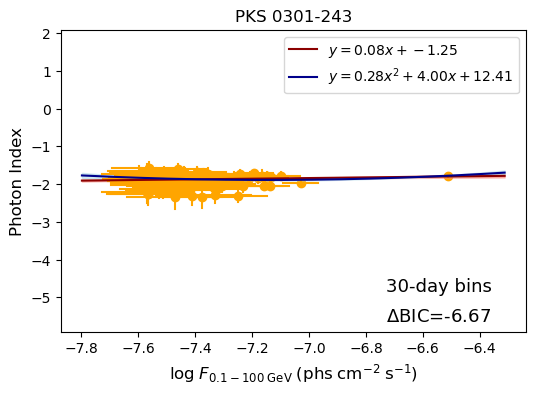}
\includegraphics[width=0.267\linewidth]{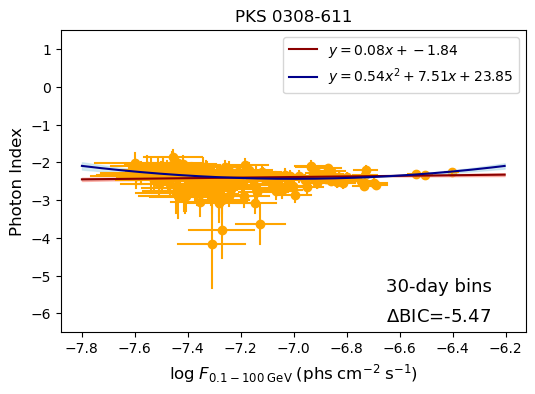}
\includegraphics[width=0.267\linewidth]{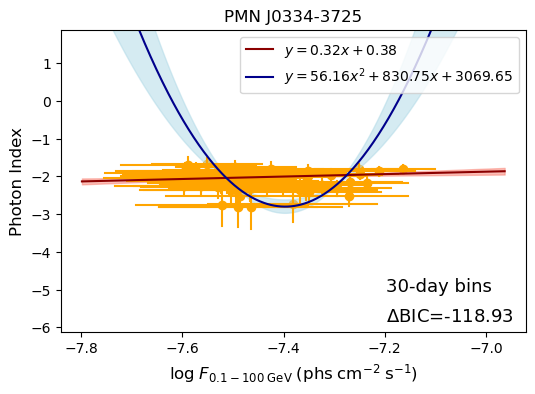}
\includegraphics[width=0.267\linewidth]{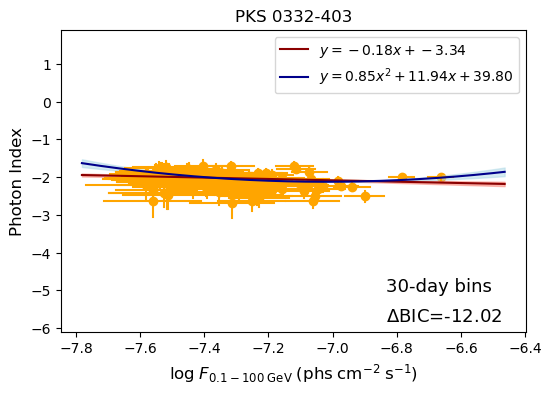}
\includegraphics[width=0.267\linewidth]{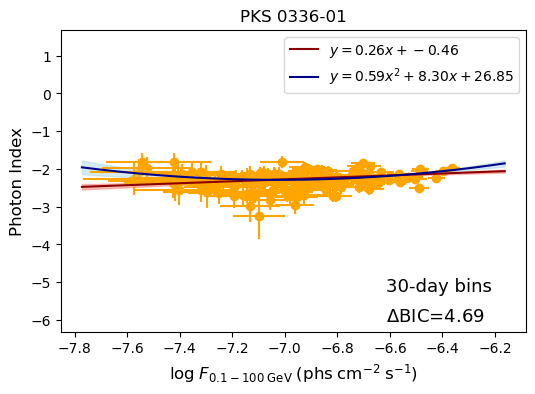}
\includegraphics[width=0.267\linewidth]{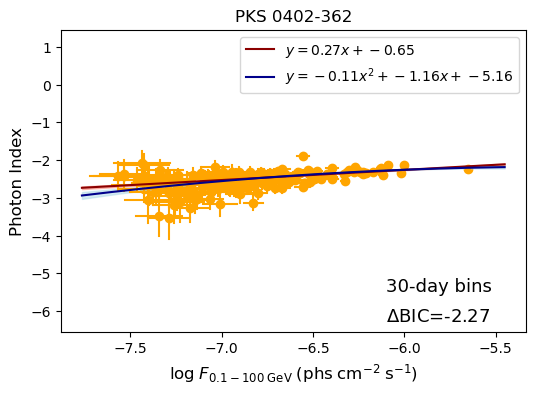}
\includegraphics[width=0.267\linewidth]{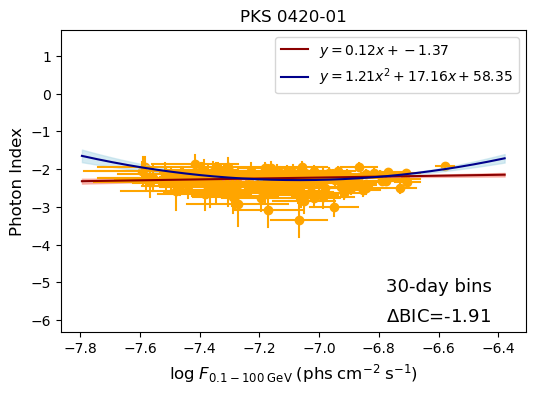}
\includegraphics[width=0.267\linewidth]{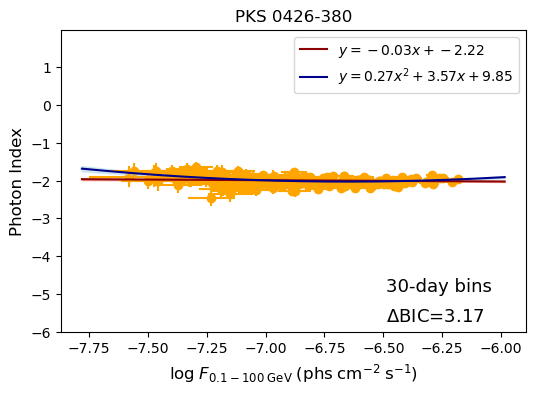}
\includegraphics[width=0.267\linewidth]{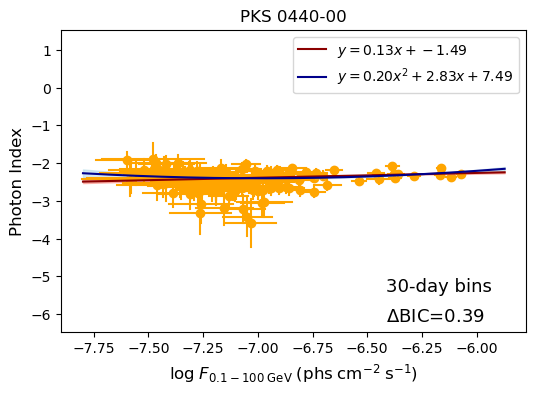}
\includegraphics[width=0.267\linewidth]{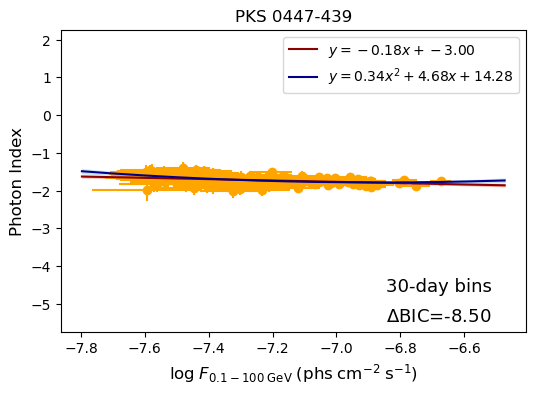}
\includegraphics[width=0.267\linewidth]{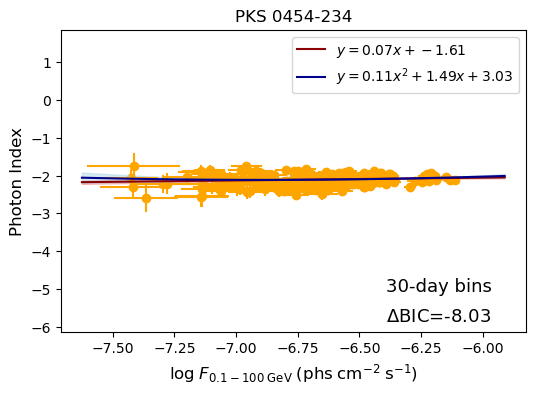}
\includegraphics[width=0.267\linewidth]{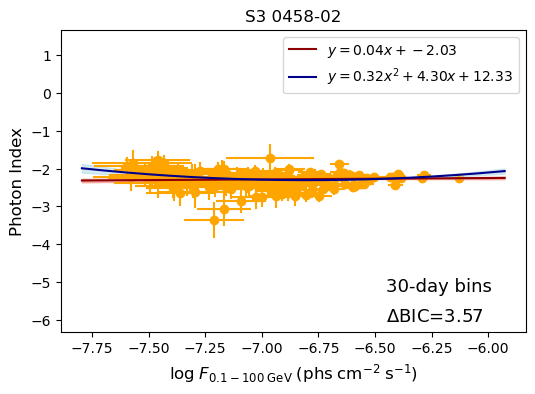}
\includegraphics[width=0.267\linewidth]{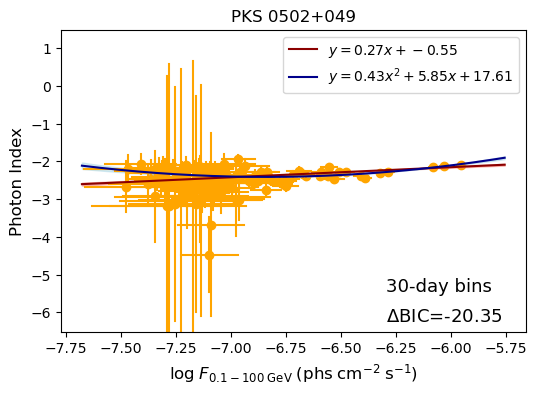}
\includegraphics[width=0.267\linewidth]{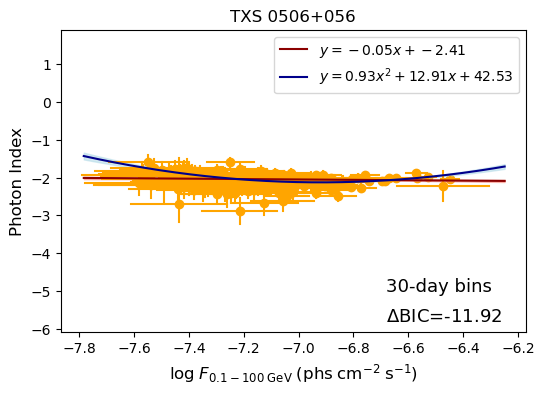}
\includegraphics[width=0.267\linewidth]{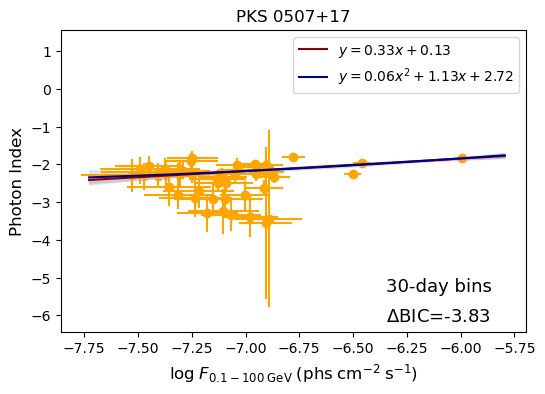}
\includegraphics[width=0.267\linewidth]{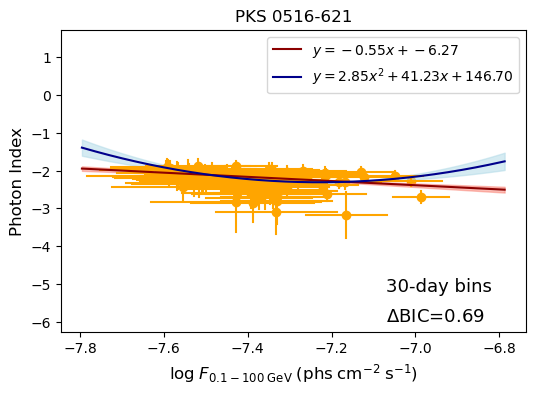}
    \caption{The same as in Figure \ref{fig:indexplot}.} 
    \label{fig:a2}
\end{figure}

\begin{figure}[h!]
	\centering
\includegraphics[width=0.267\linewidth]{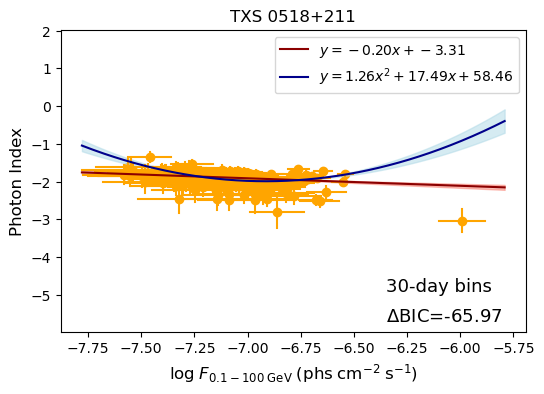}
\includegraphics[width=0.267\linewidth]{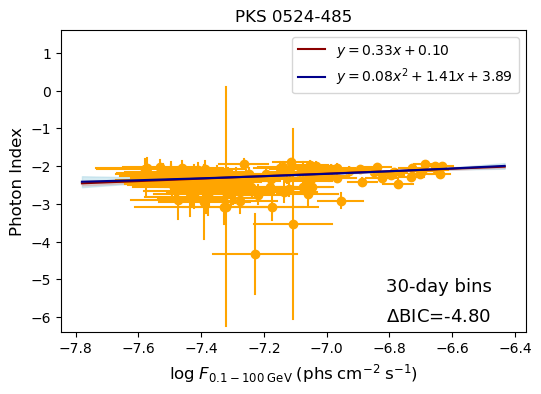}
\includegraphics[width=0.267\linewidth]{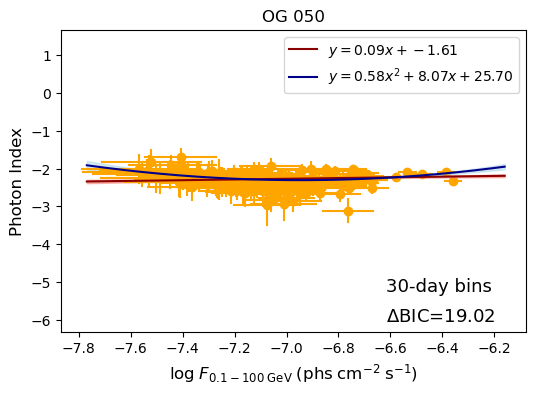}
\includegraphics[width=0.267\linewidth]{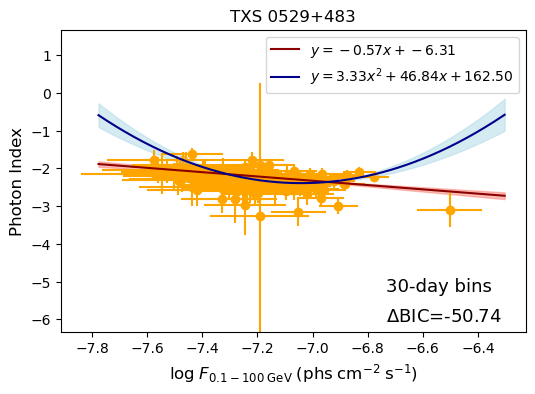}
\includegraphics[width=0.267\linewidth]{PKS_0537-441.png}
\includegraphics[width=0.267\linewidth]{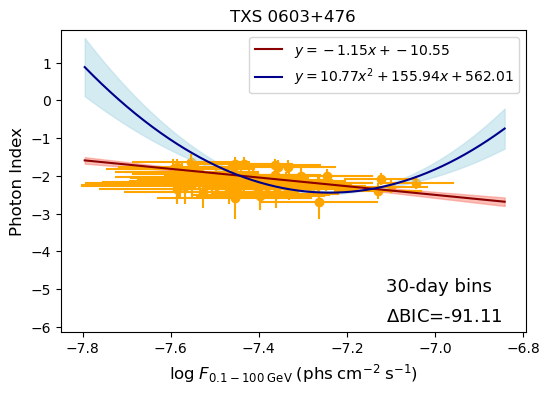}
\includegraphics[width=0.267\linewidth]{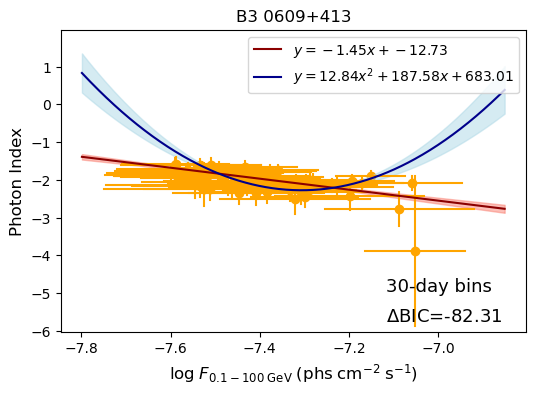}
\includegraphics[width=0.267\linewidth]{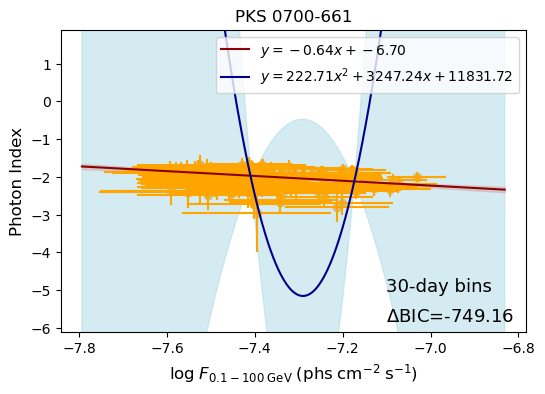}
\includegraphics[width=0.267\linewidth]{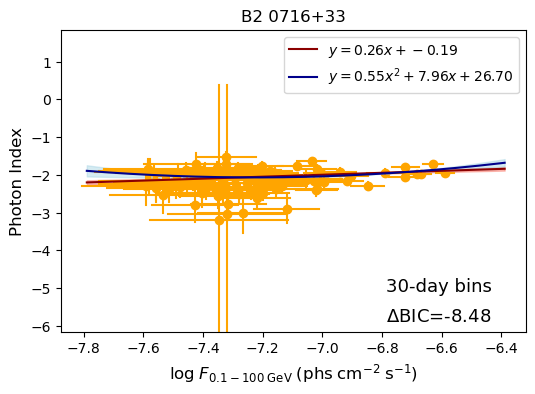}
\includegraphics[width=0.267\linewidth]{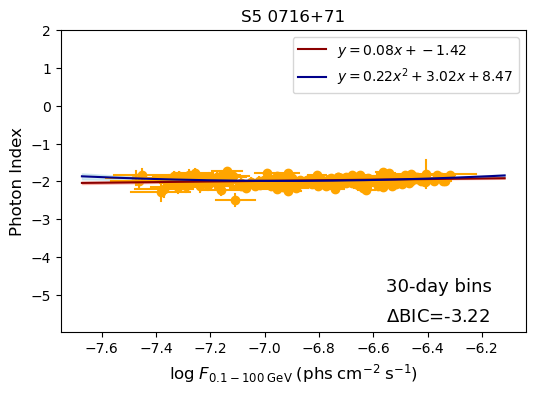}
\includegraphics[width=0.267\linewidth]{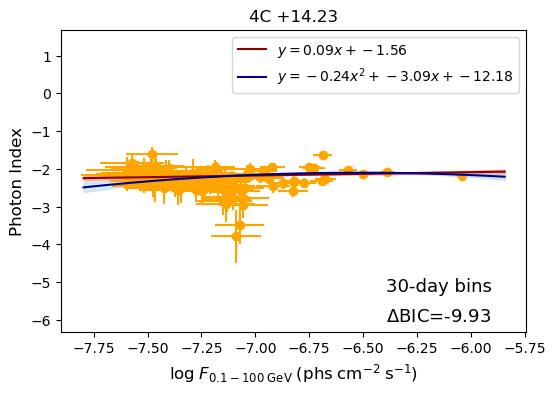}
\includegraphics[width=0.267\linewidth]{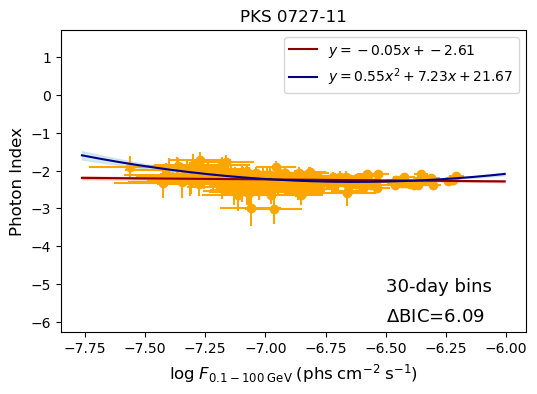}
\includegraphics[width=0.267\linewidth]{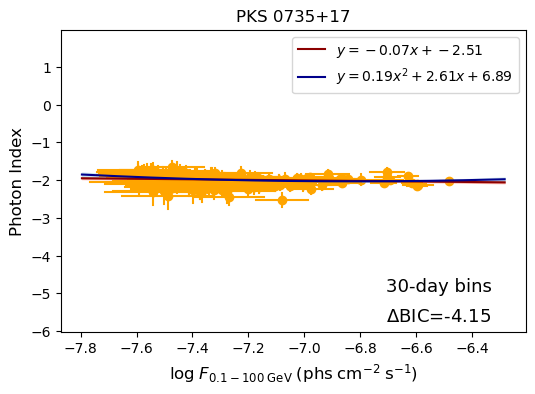}
\includegraphics[width=0.267\linewidth]{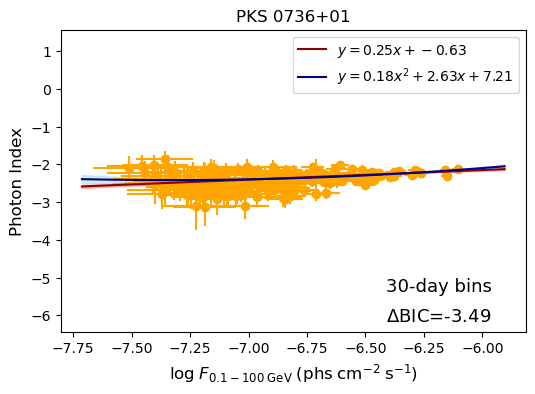}
\includegraphics[width=0.267\linewidth]{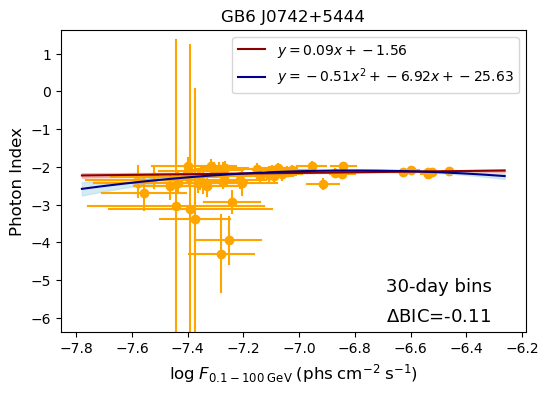}
\includegraphics[width=0.267\linewidth]{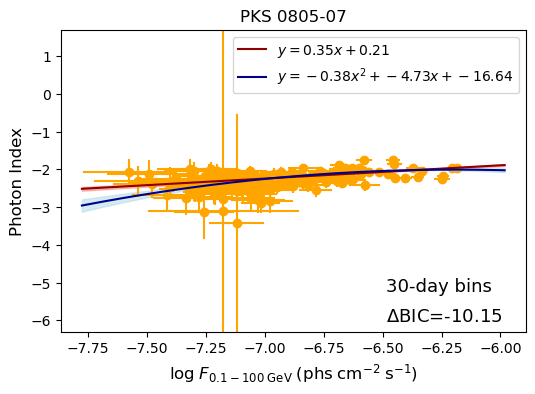}
\includegraphics[width=0.267\linewidth]{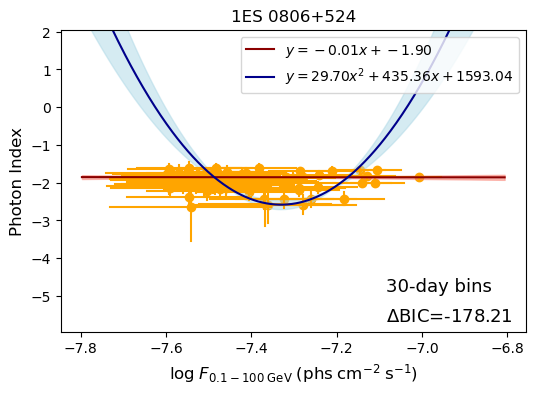}
\includegraphics[width=0.267\linewidth]{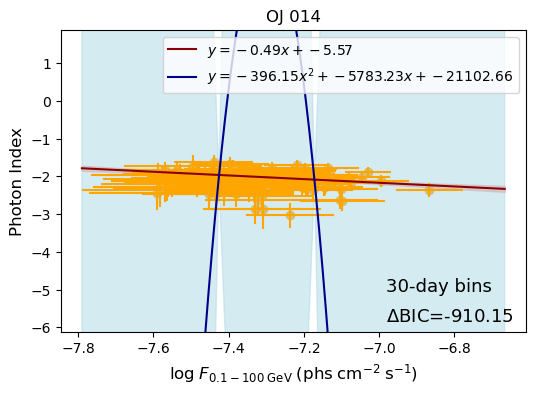}
    \caption{The same as in Figure \ref{fig:indexplot}.} 
    \label{fig:a3}
\end{figure}

\begin{figure}[h!]
	\centering
\includegraphics[width=0.267\linewidth]{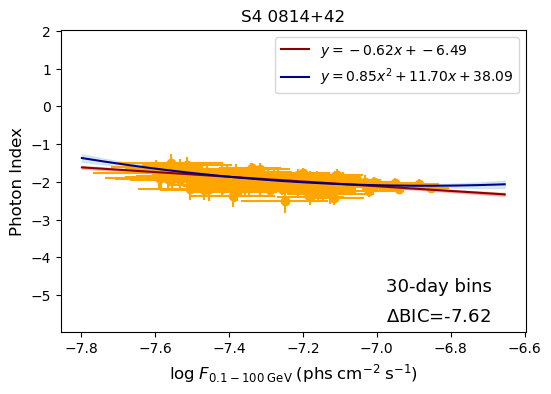}
\includegraphics[width=0.267\linewidth]{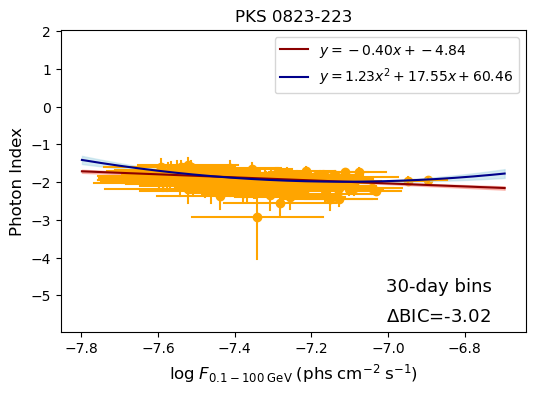}
\includegraphics[width=0.267\linewidth]{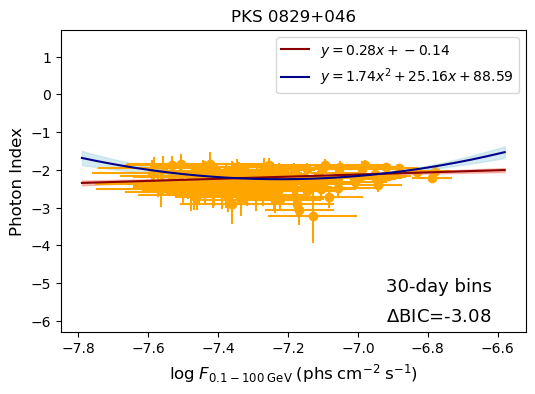}
\includegraphics[width=0.267\linewidth]{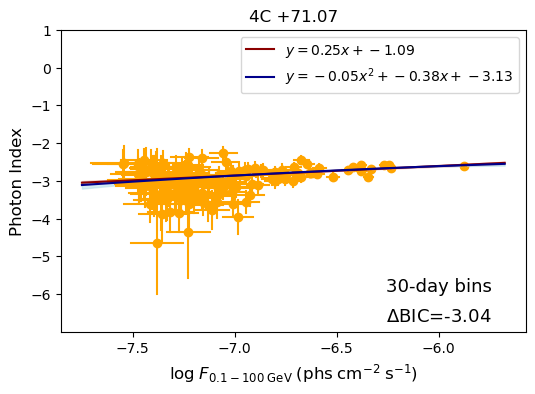}
\includegraphics[width=0.267\linewidth]{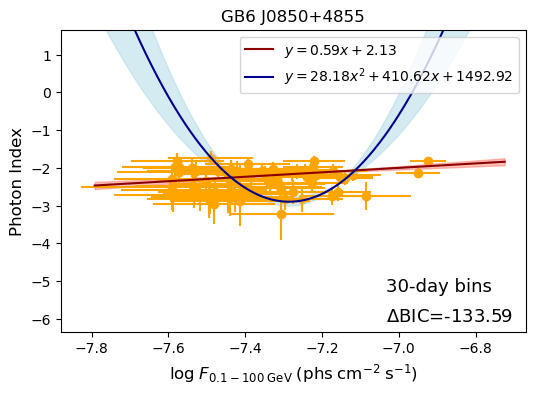}
\includegraphics[width=0.267\linewidth]{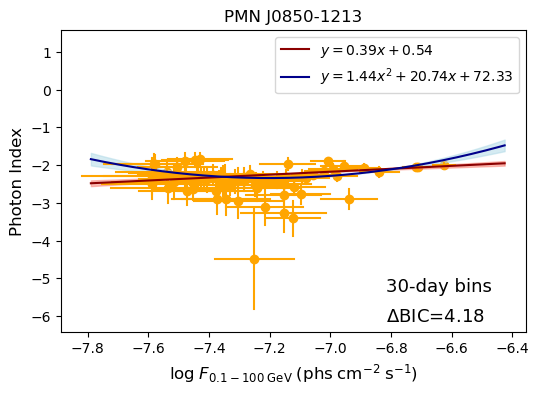}
\includegraphics[width=0.267\linewidth]{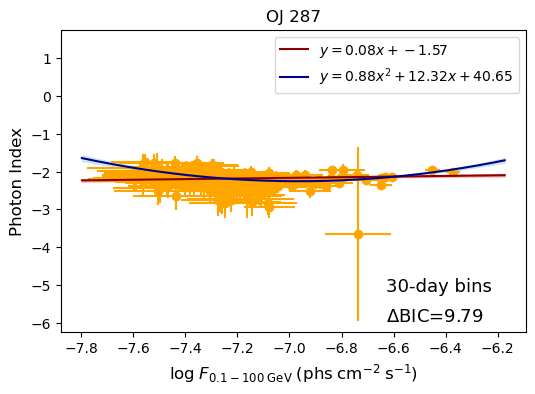}
\includegraphics[width=0.267\linewidth]{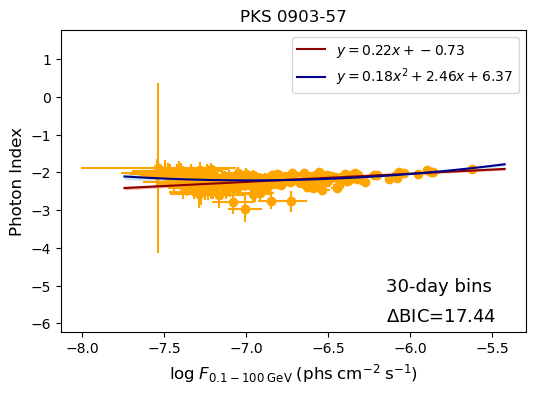}
\includegraphics[width=0.267\linewidth]{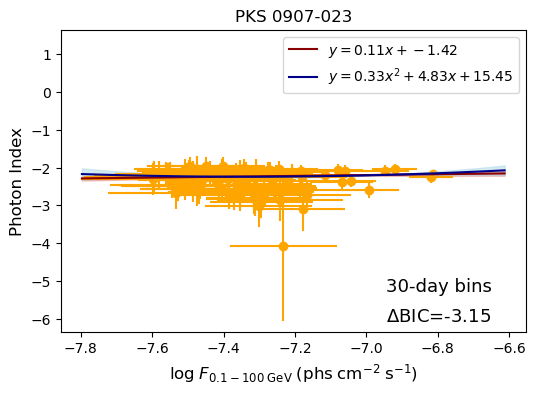}
\includegraphics[width=0.267\linewidth]{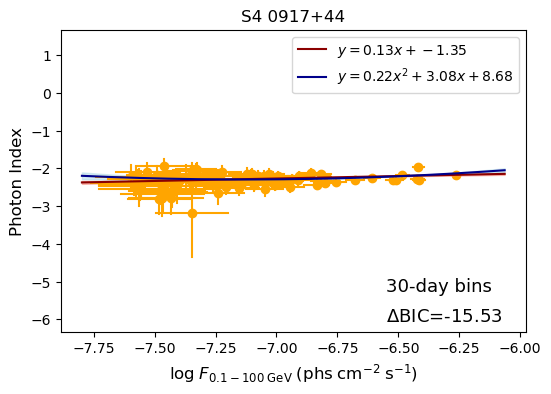}
\includegraphics[width=0.267\linewidth]{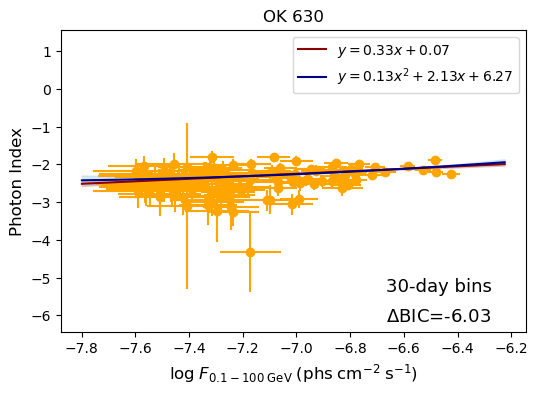}
\includegraphics[width=0.267\linewidth]{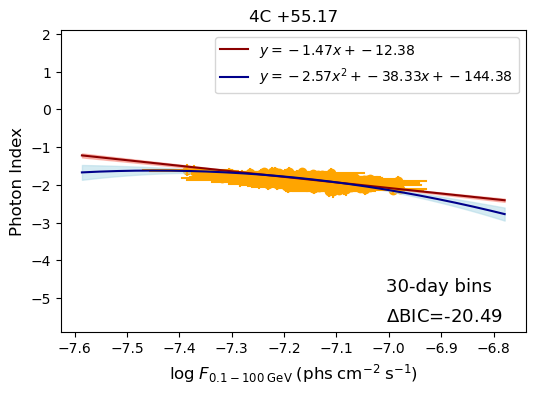}
\includegraphics[width=0.267\linewidth]{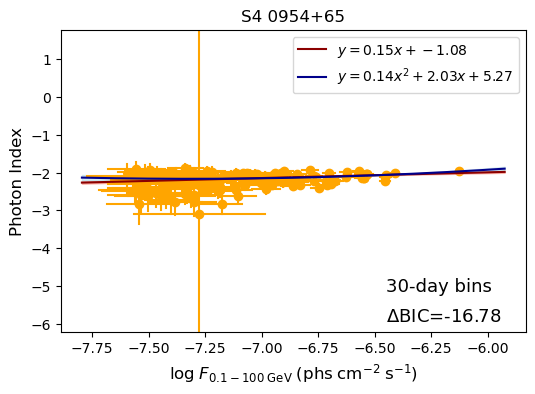}
\includegraphics[width=0.267\linewidth]{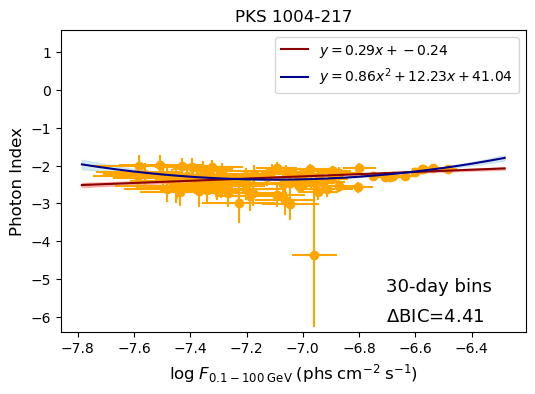}
\includegraphics[width=0.267\linewidth]{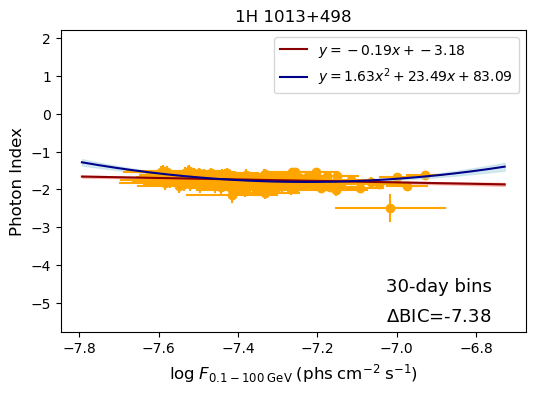}
\includegraphics[width=0.267\linewidth]{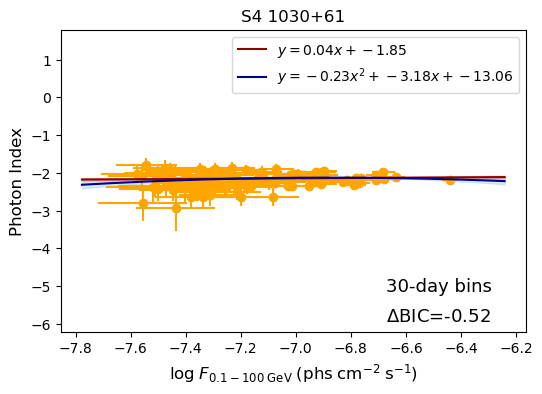}
\includegraphics[width=0.267\linewidth]{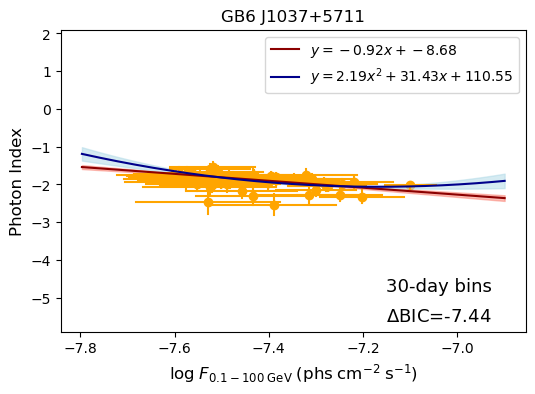}
\includegraphics[width=0.267\linewidth]{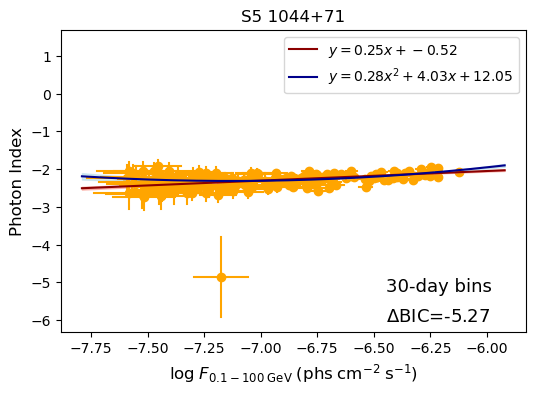}
    \caption{The same as in Figure \ref{fig:indexplot}.} 
    \label{fig:a4}
\end{figure}

\begin{figure}[h!]
	\centering
\includegraphics[width=0.267\linewidth]{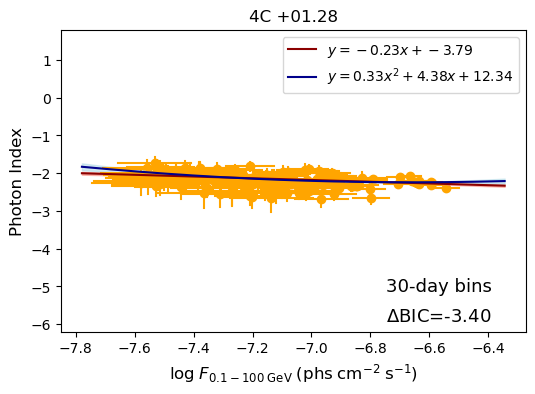}
\includegraphics[width=0.267\linewidth]{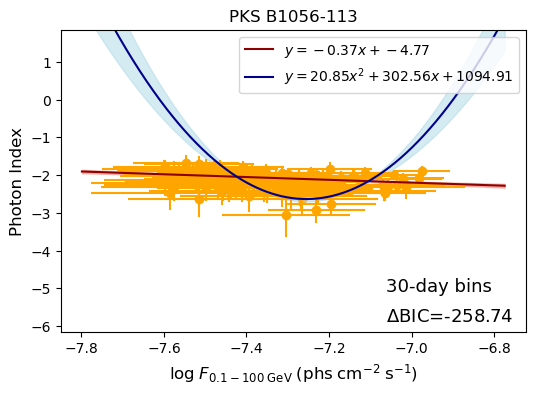}
\includegraphics[width=0.267\linewidth]{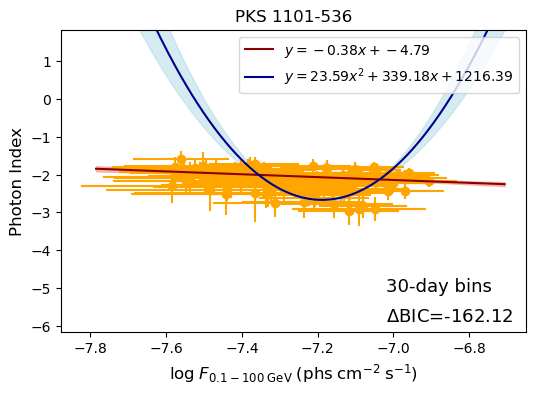}
\includegraphics[width=0.267\linewidth]{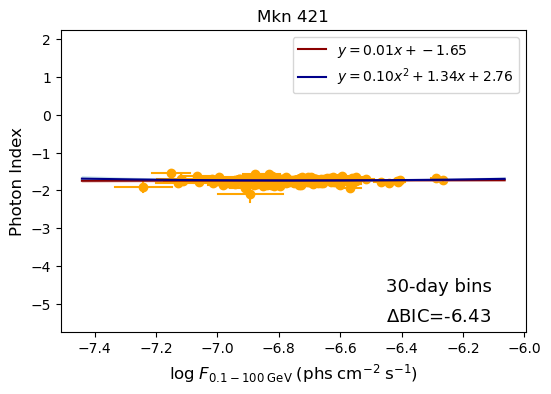}
\includegraphics[width=0.267\linewidth]{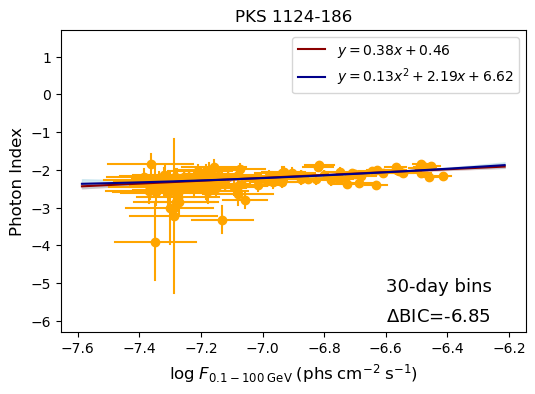}
\includegraphics[width=0.267\linewidth]{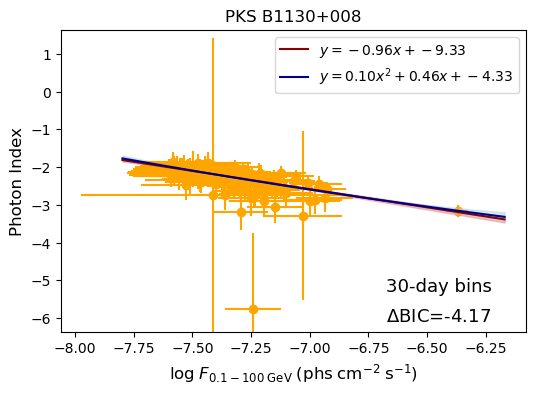}
\includegraphics[width=0.267\linewidth]{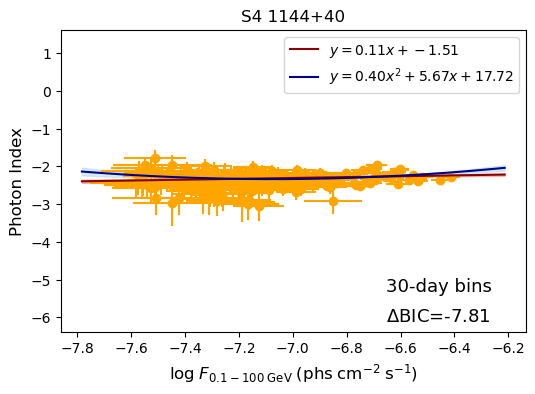}
\includegraphics[width=0.267\linewidth]{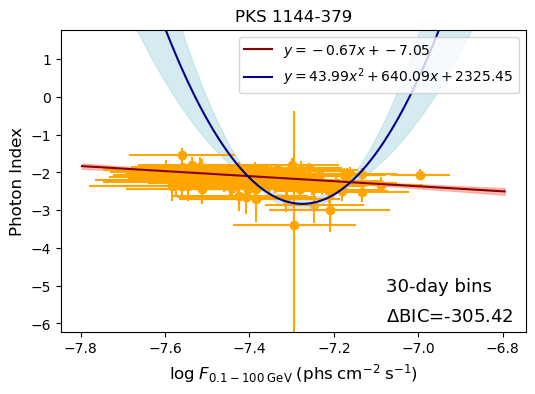}
\includegraphics[width=0.267\linewidth]{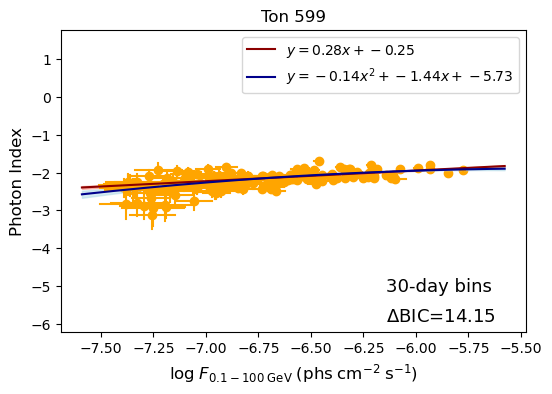}
\includegraphics[width=0.267\linewidth]{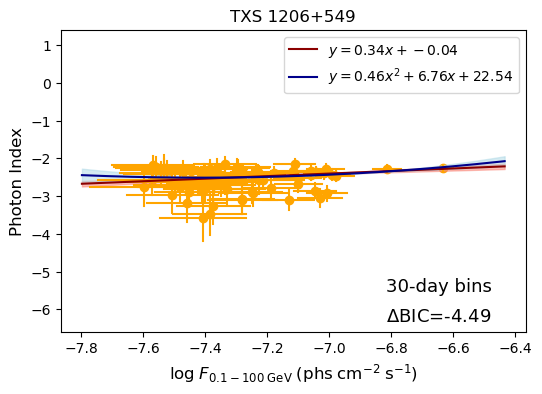}
\includegraphics[width=0.267\linewidth]{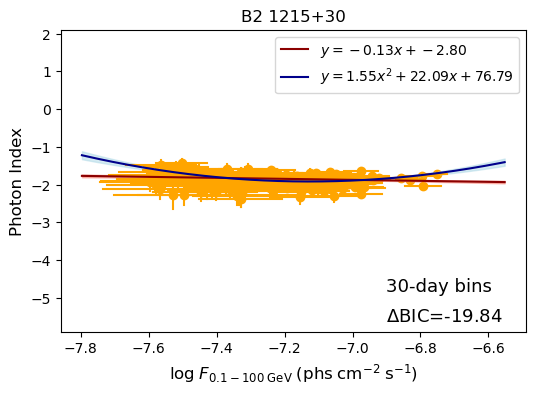}
\includegraphics[width=0.267\linewidth]{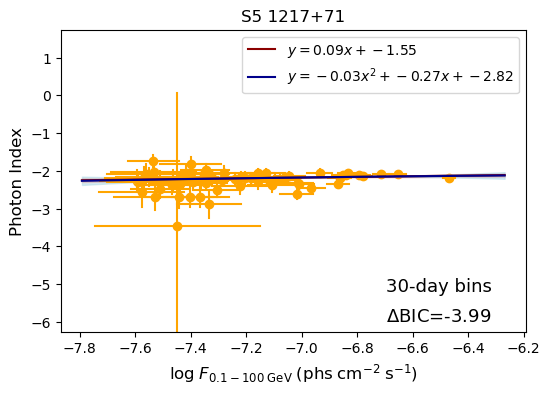}
\includegraphics[width=0.267\linewidth]{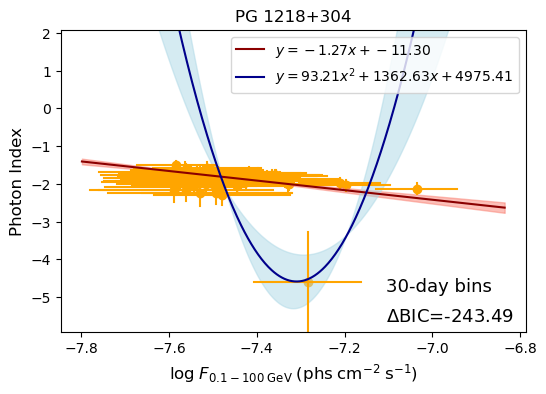}
\includegraphics[width=0.267\linewidth]{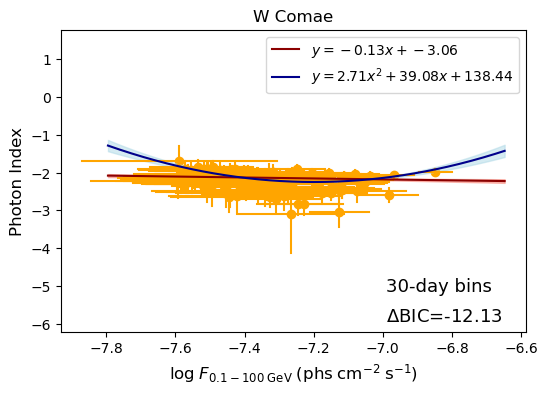}
\includegraphics[width=0.267\linewidth]{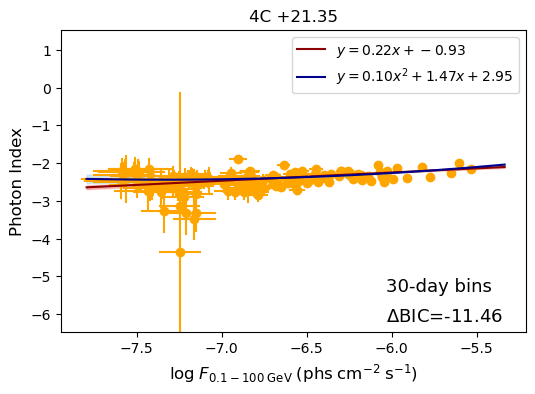}
\includegraphics[width=0.267\linewidth]{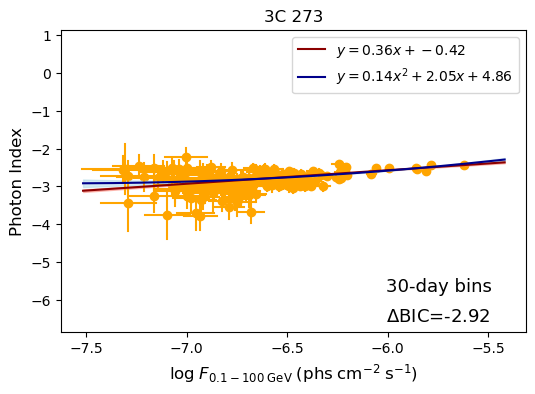}
\includegraphics[width=0.267\linewidth]{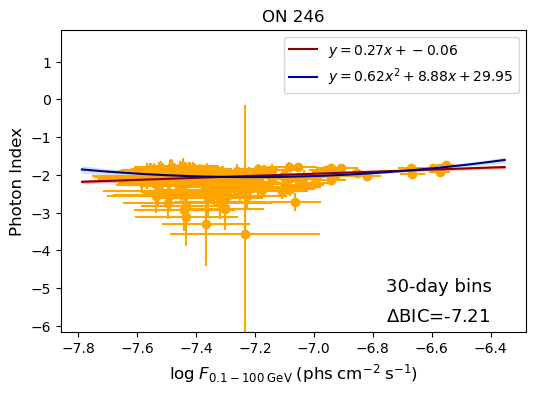}
\includegraphics[width=0.267\linewidth]{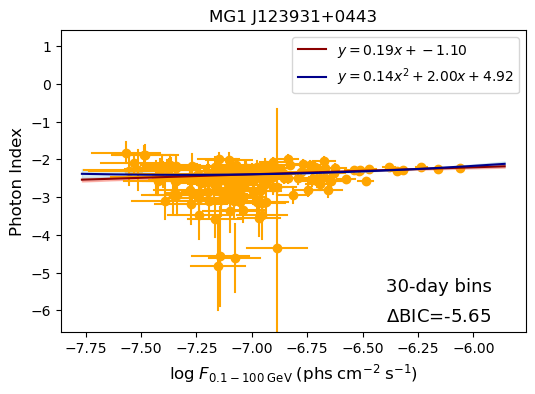}
    \caption{The same as in Figure \ref{fig:indexplot}.} 
    \label{fig:a5}
\end{figure}

\begin{figure}[h!]
	\centering
\includegraphics[width=0.267\linewidth]{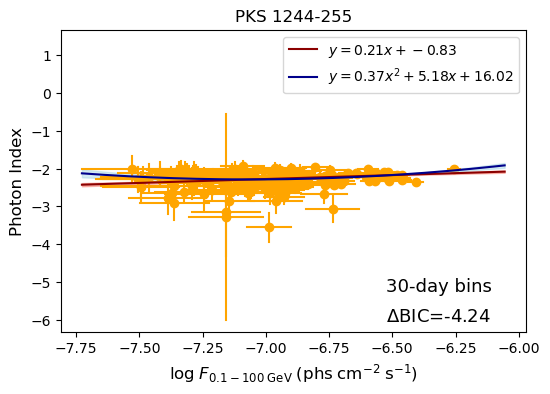}
\includegraphics[width=0.267\linewidth]{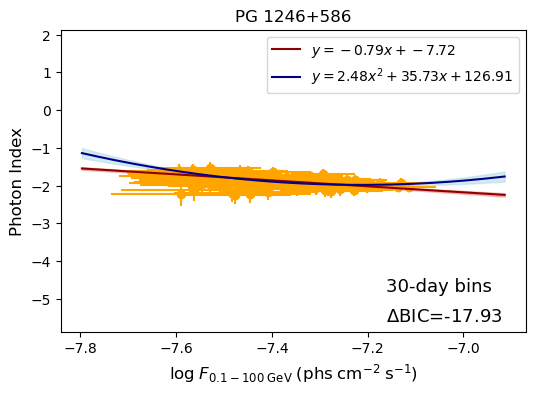}
\includegraphics[width=0.267\linewidth]{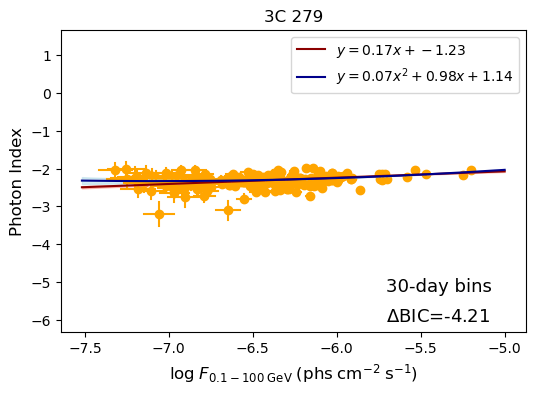}
\includegraphics[width=0.267\linewidth]{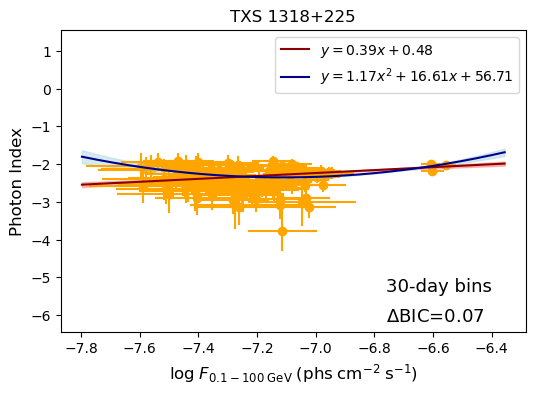}
\includegraphics[width=0.267\linewidth]{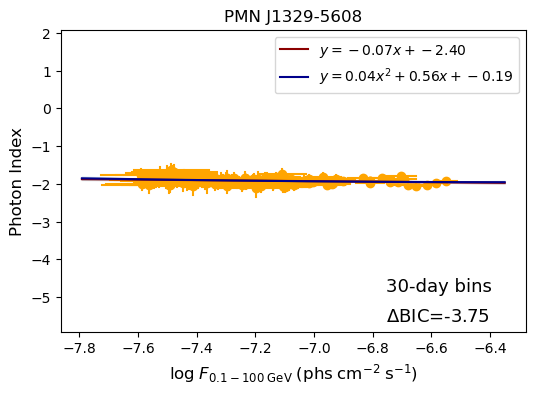}
\includegraphics[width=0.267\linewidth]{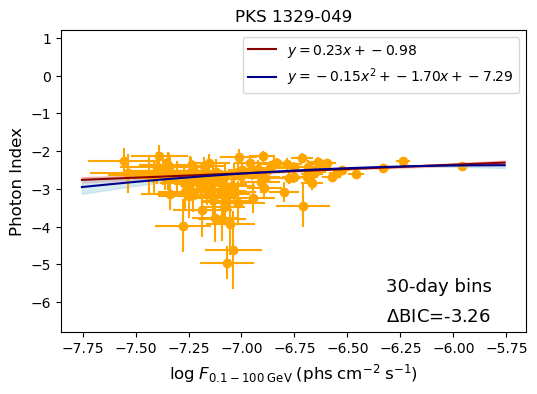}
\includegraphics[width=0.267\linewidth]{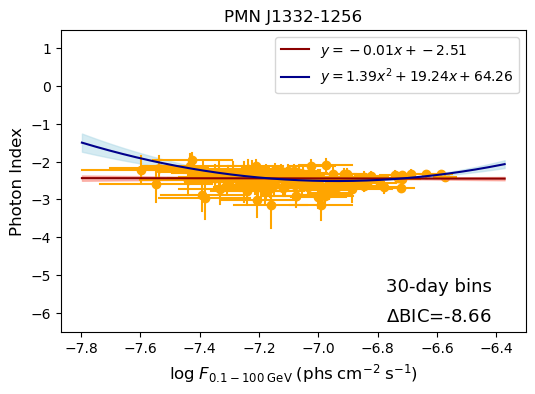}
\includegraphics[width=0.267\linewidth]{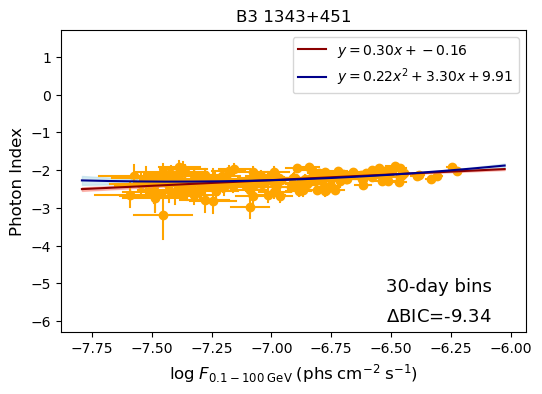}
\includegraphics[width=0.267\linewidth]{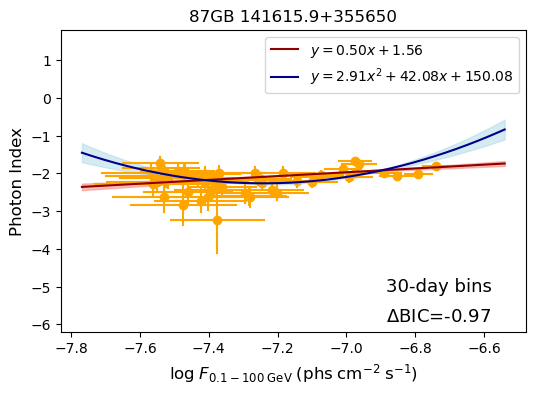}
\includegraphics[width=0.267\linewidth]{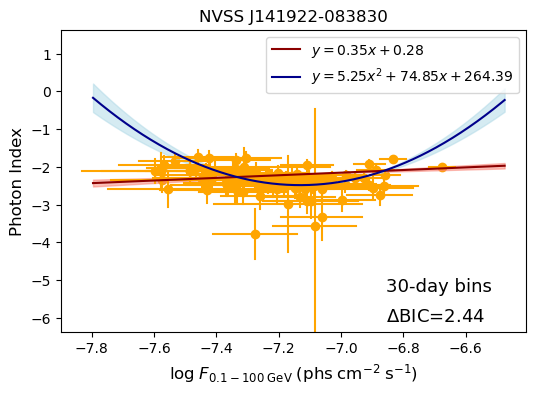}
\includegraphics[width=0.267\linewidth]{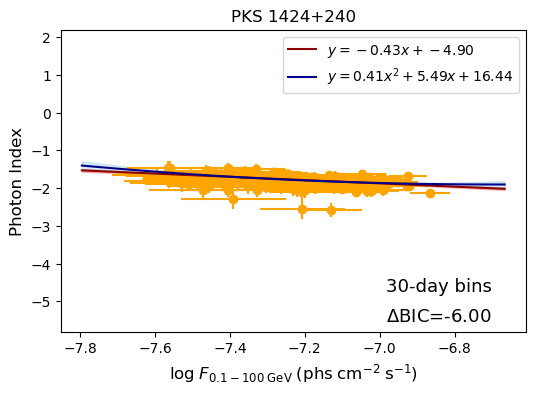}
\includegraphics[width=0.267\linewidth]{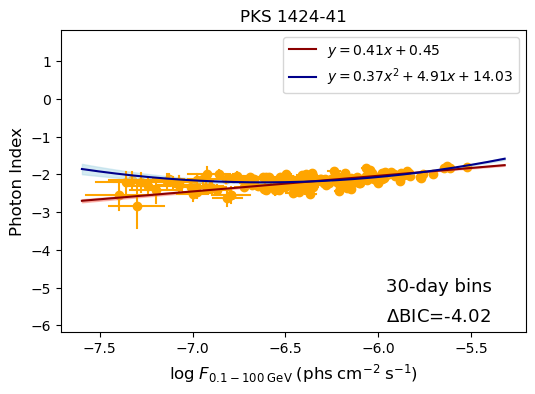}
\includegraphics[width=0.267\linewidth]{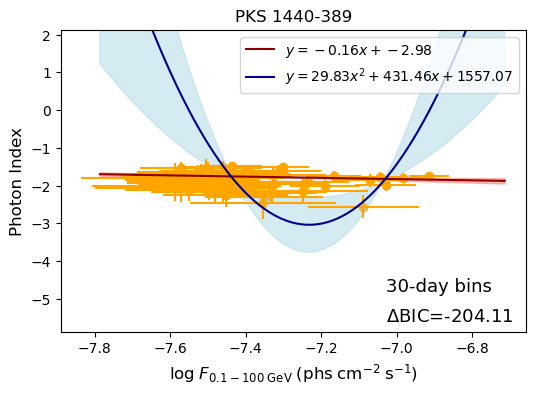}
\includegraphics[width=0.267\linewidth]{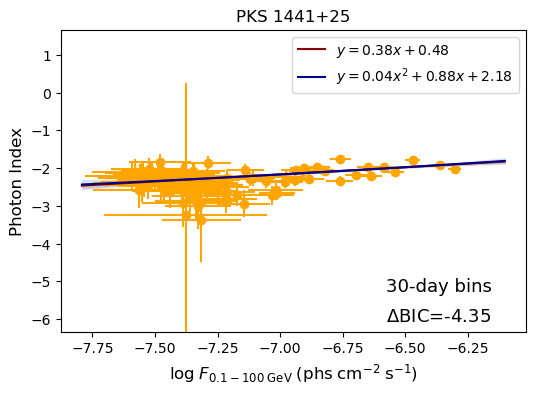}
\includegraphics[width=0.267\linewidth]{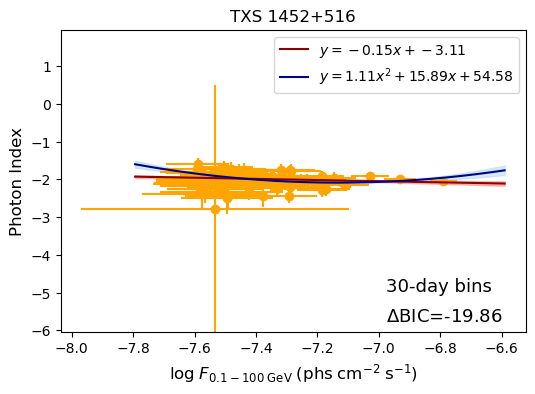}
\includegraphics[width=0.267\linewidth]{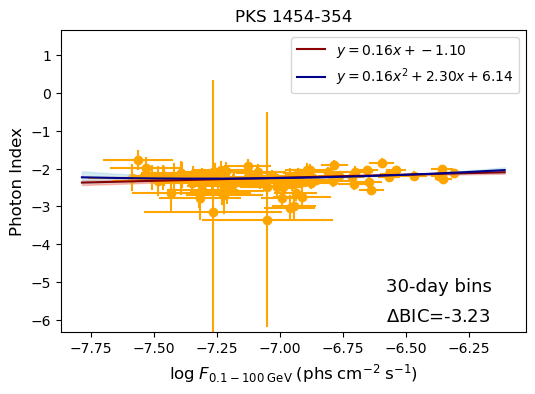}
\includegraphics[width=0.267\linewidth]{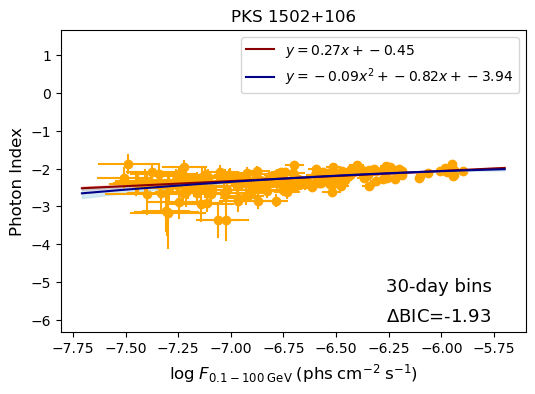}
\includegraphics[width=0.267\linewidth]{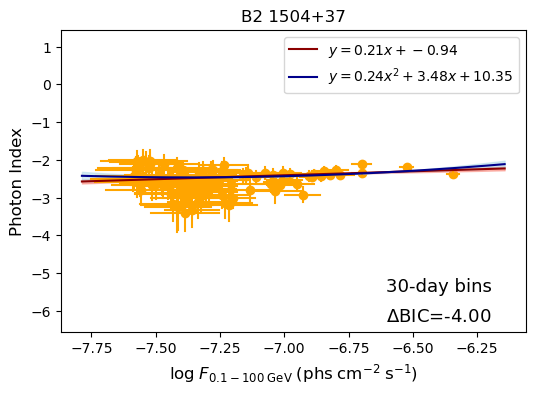}
    \caption{The same as in Figure \ref{fig:indexplot}.} 
    \label{fig:a6}
\end{figure}

\begin{figure}[h!]
	\centering
\includegraphics[width=0.267\linewidth]{PKS_1510-089.png}
\includegraphics[width=0.267\linewidth]{AP_Librae.png}
\includegraphics[width=0.267\linewidth]{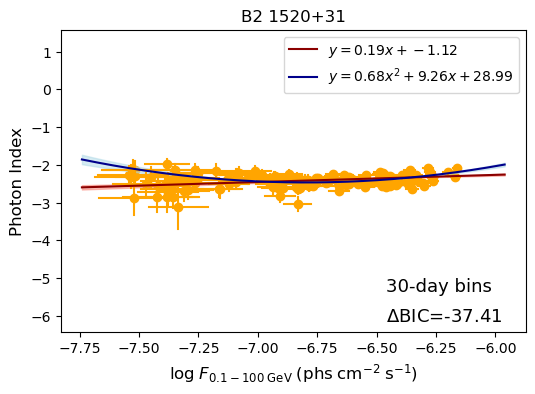}
\includegraphics[width=0.267\linewidth]{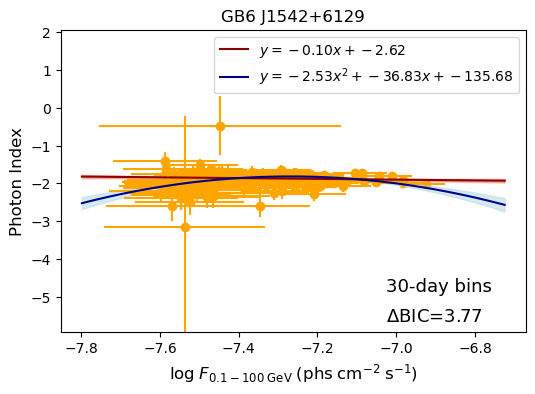}
\includegraphics[width=0.267\linewidth]{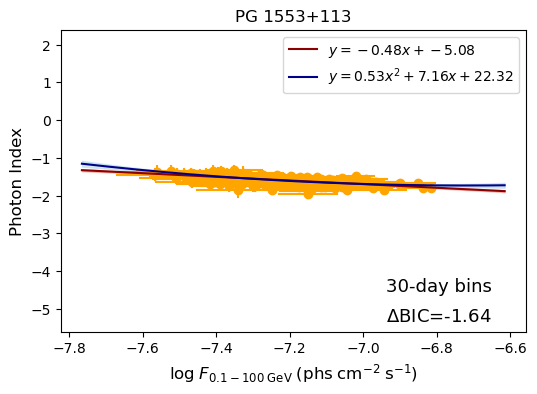}
\includegraphics[width=0.267\linewidth]{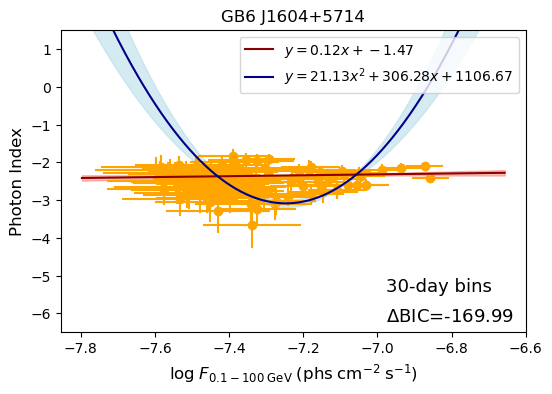}
\includegraphics[width=0.267\linewidth]{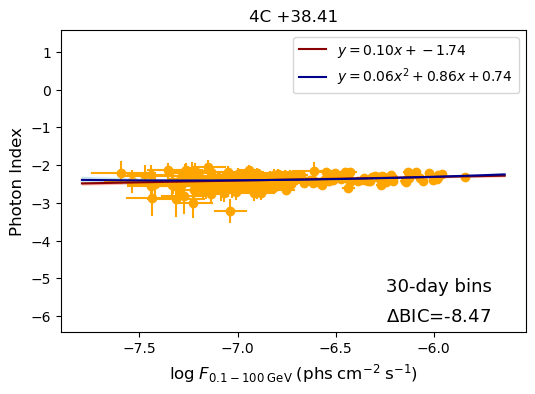}
\includegraphics[width=0.267\linewidth]{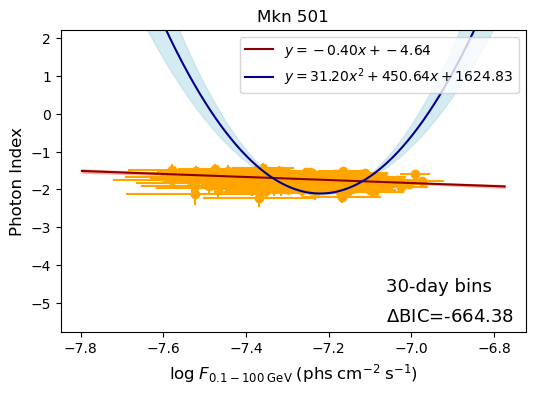}
\includegraphics[width=0.267\linewidth]{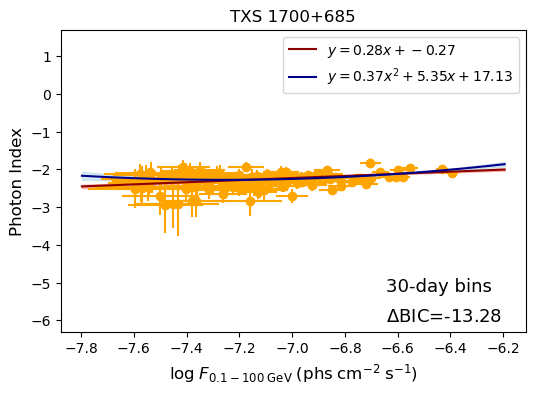}
\includegraphics[width=0.267\linewidth]{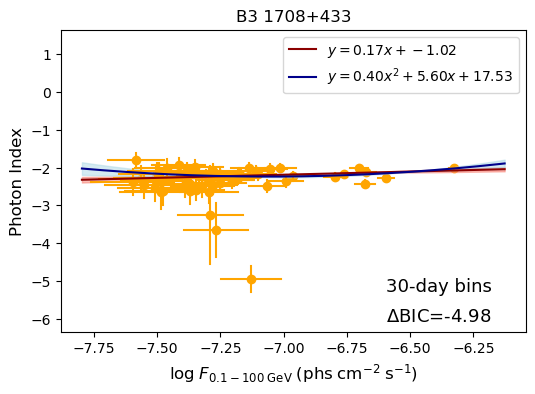}
\includegraphics[width=0.267\linewidth]{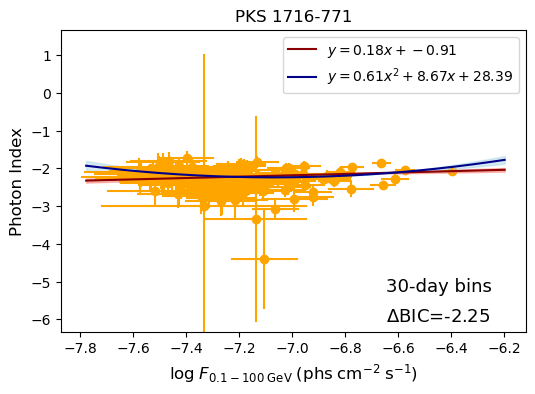}
\includegraphics[width=0.267\linewidth]{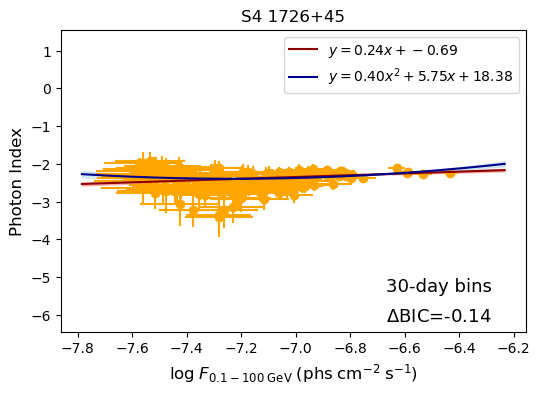}
\includegraphics[width=0.267\linewidth]{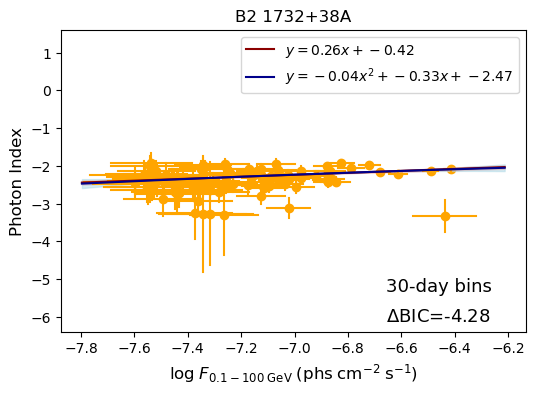}
\includegraphics[width=0.267\linewidth]{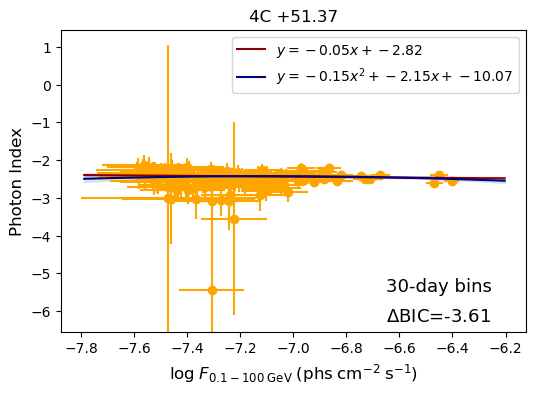}
\includegraphics[width=0.267\linewidth]{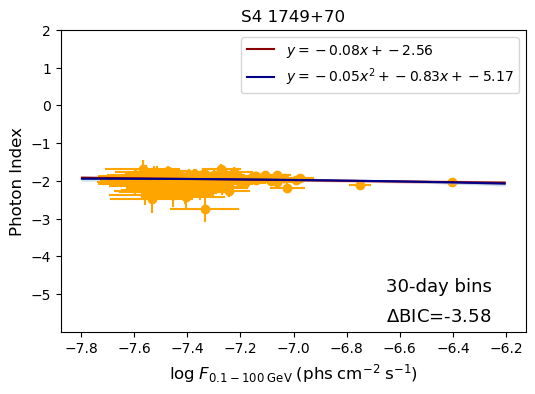}
\includegraphics[width=0.267\linewidth]{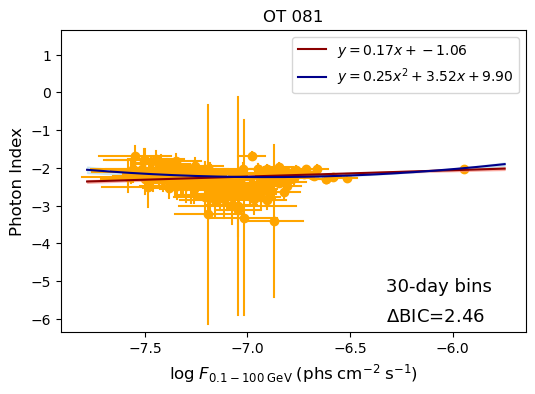}
\includegraphics[width=0.267\linewidth]{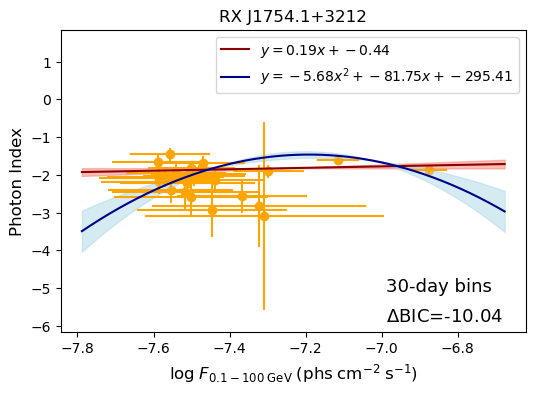}
\includegraphics[width=0.267\linewidth]{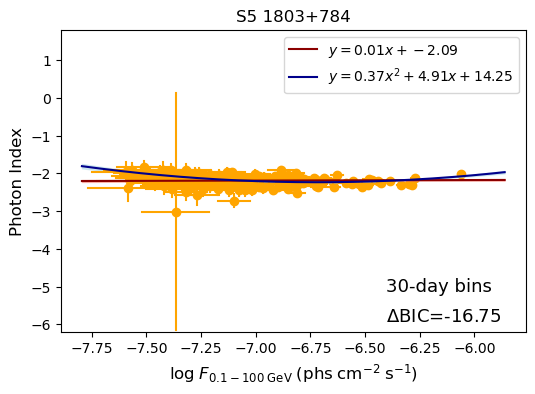}
    \caption{The same as in Figure \ref{fig:indexplot}.} 
    \label{fig:a7}
\end{figure}

\begin{figure}[h!]
	\centering
\includegraphics[width=0.267\linewidth]{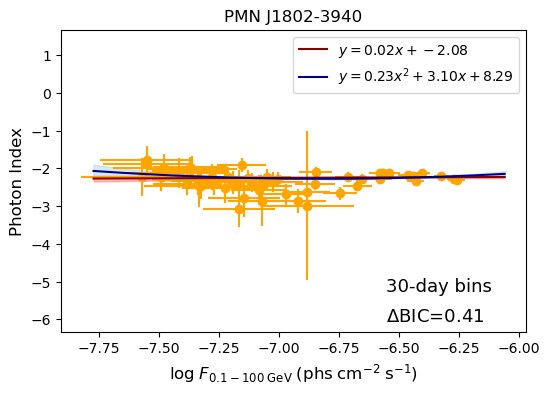}
\includegraphics[width=0.267\linewidth]{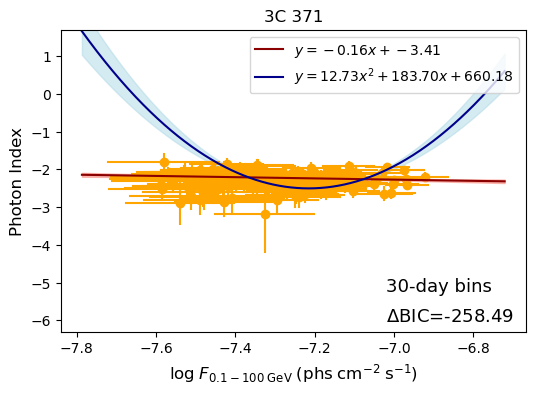}
\includegraphics[width=0.267\linewidth]{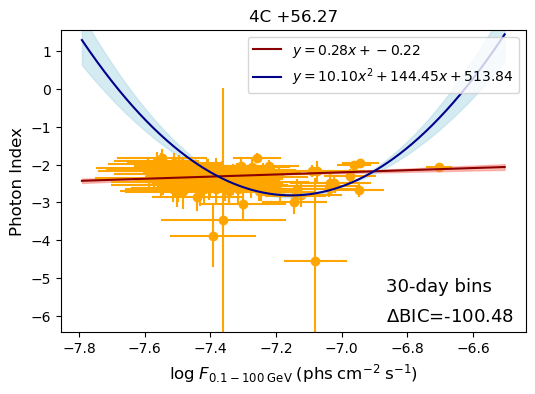}
\includegraphics[width=0.267\linewidth]{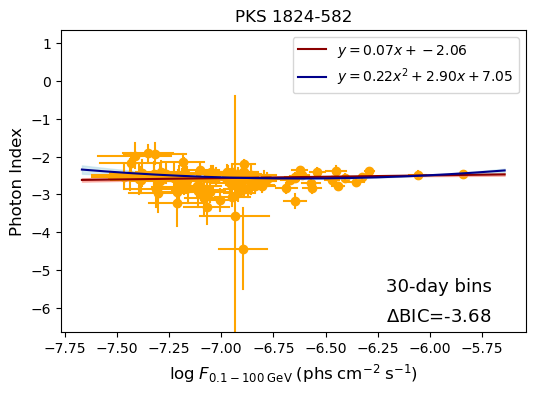}
\includegraphics[width=0.267\linewidth]{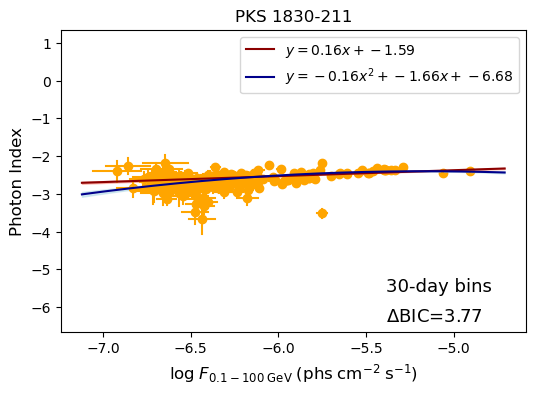}
\includegraphics[width=0.267\linewidth]{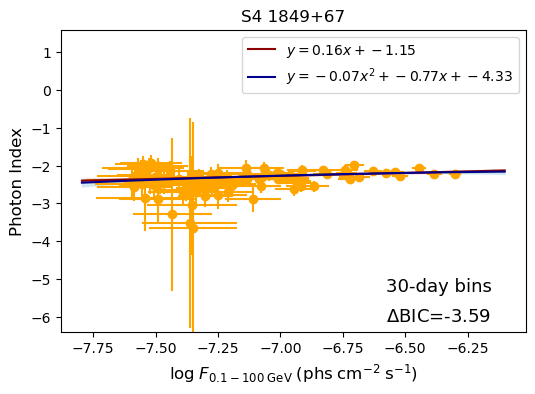}
\includegraphics[width=0.267\linewidth]{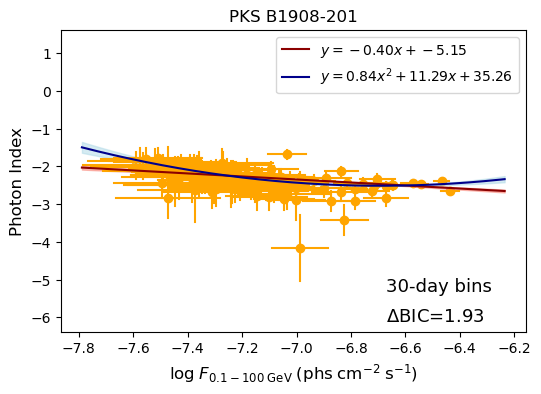}
\includegraphics[width=0.267\linewidth]{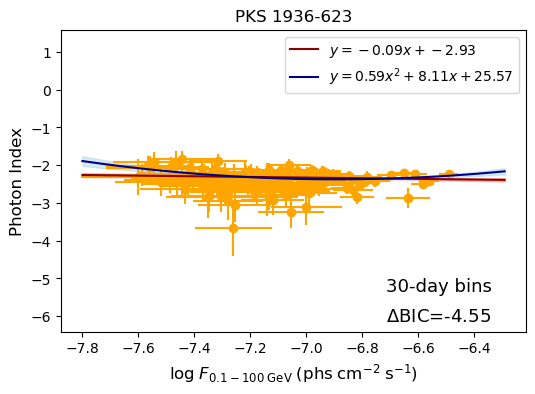}
\includegraphics[width=0.267\linewidth]{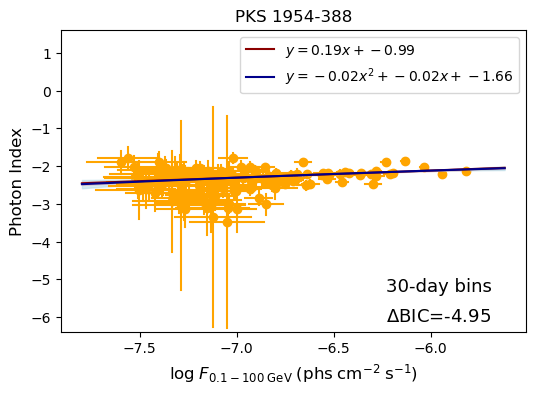}
\includegraphics[width=0.267\linewidth]{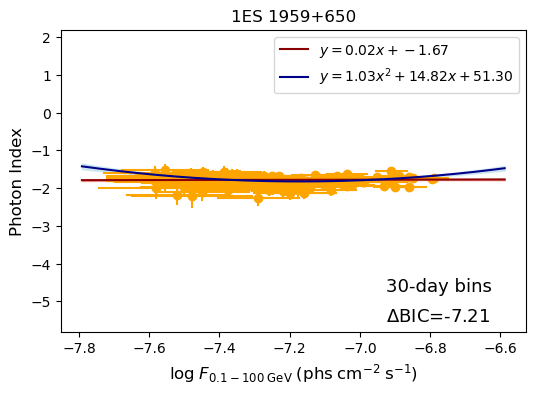}
\includegraphics[width=0.267\linewidth]{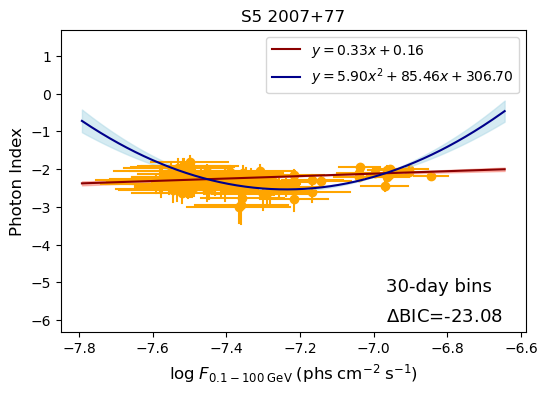}
\includegraphics[width=0.267\linewidth]{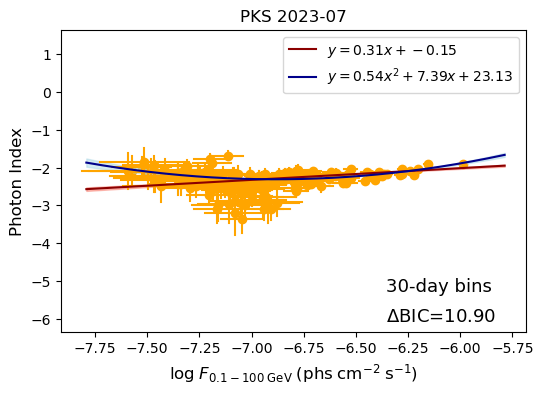}
\includegraphics[width=0.267\linewidth]{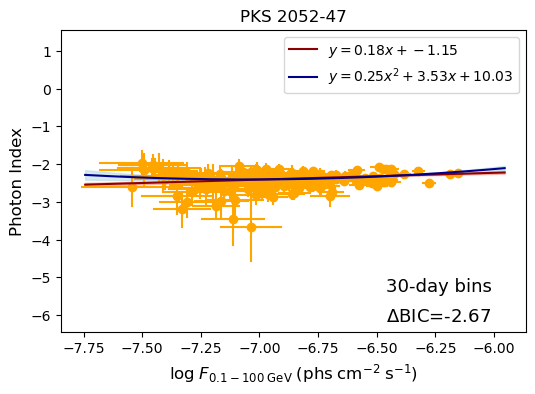}
\includegraphics[width=0.267\linewidth]{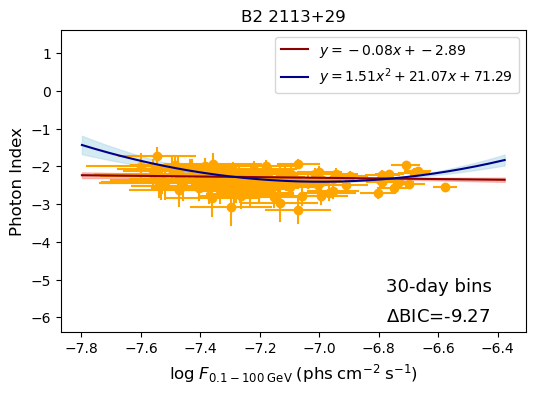}
\includegraphics[width=0.267\linewidth]{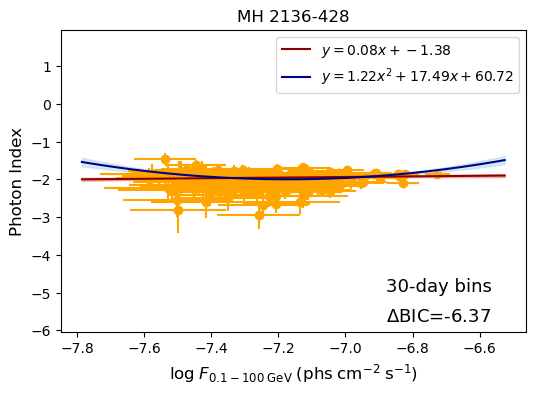}
\includegraphics[width=0.267\linewidth]{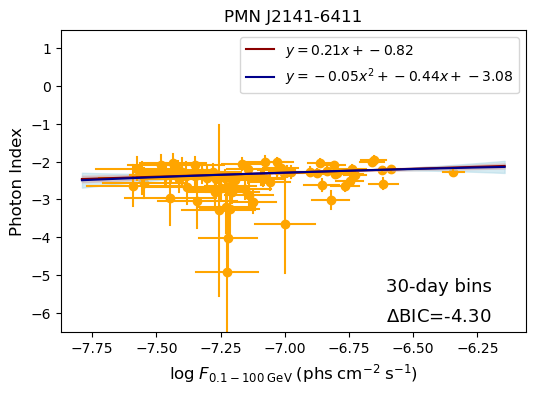}
\includegraphics[width=0.267\linewidth]{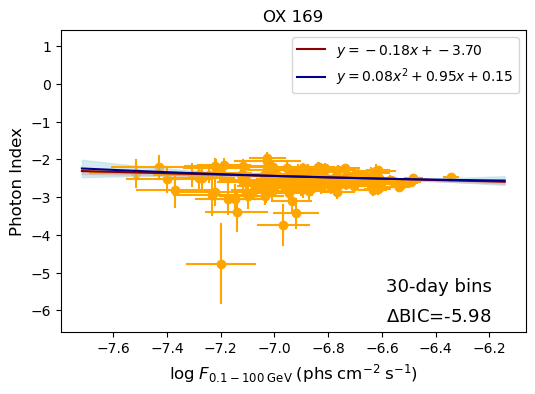}
\includegraphics[width=0.267\linewidth]{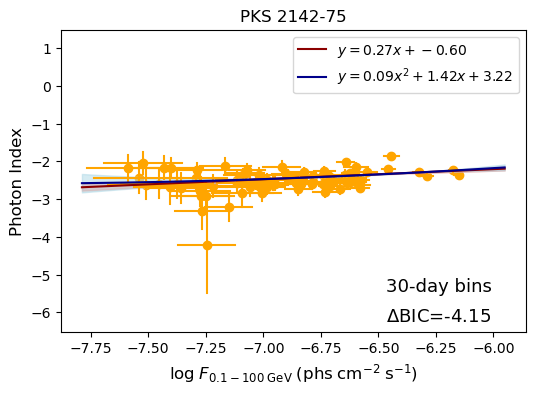}
    \caption{The same as in Figure \ref{fig:indexplot}.} 
    \label{fig:a8}
\end{figure}

\begin{figure}[h!]
	\centering
\includegraphics[width=0.267\linewidth]{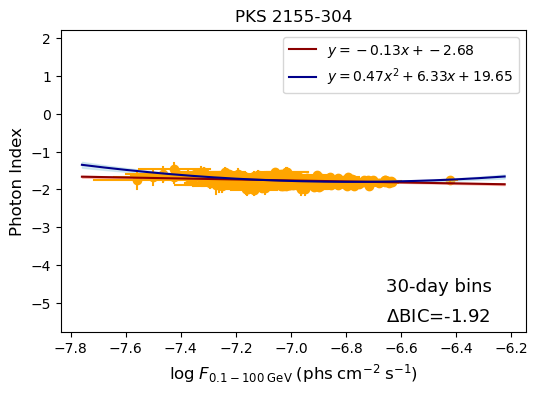}
\includegraphics[width=0.267\linewidth]{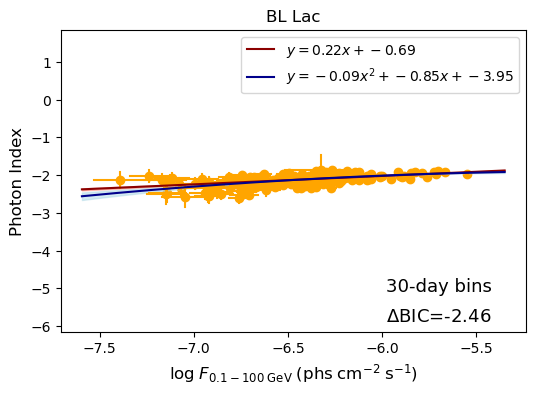}
\includegraphics[width=0.267\linewidth]{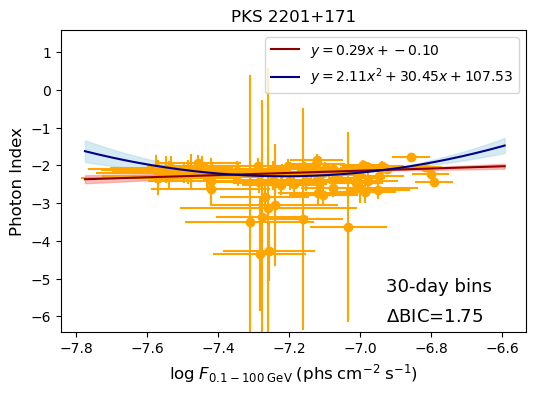}
\includegraphics[width=0.267\linewidth]{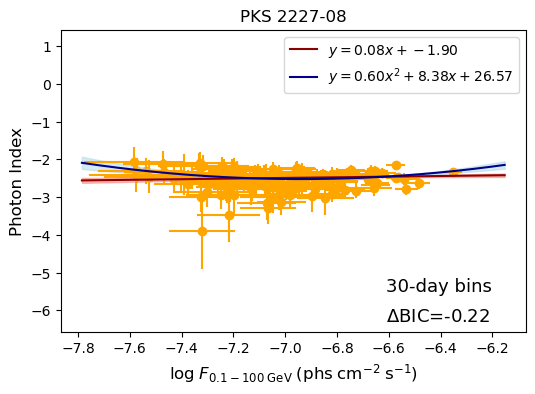}
\includegraphics[width=0.267\linewidth]{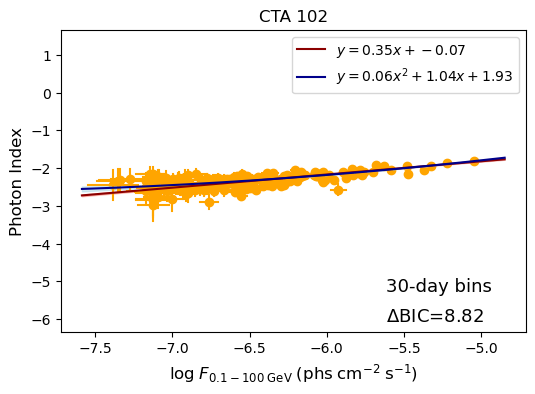}
\includegraphics[width=0.267\linewidth]{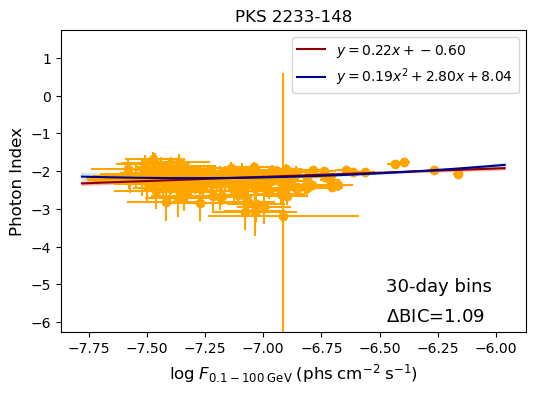}
\includegraphics[width=0.267\linewidth]{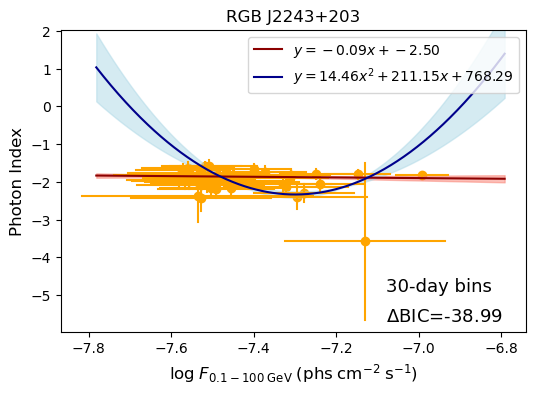}
\includegraphics[width=0.267\linewidth]{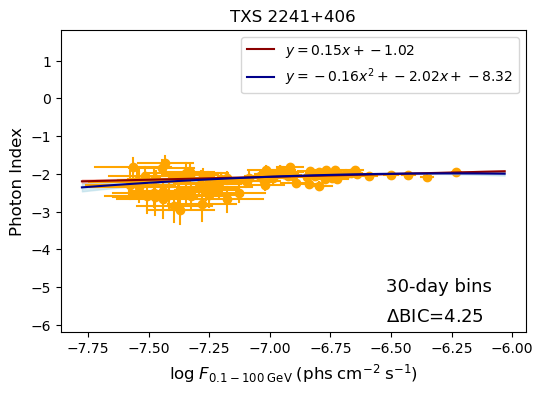}
\includegraphics[width=0.267\linewidth]{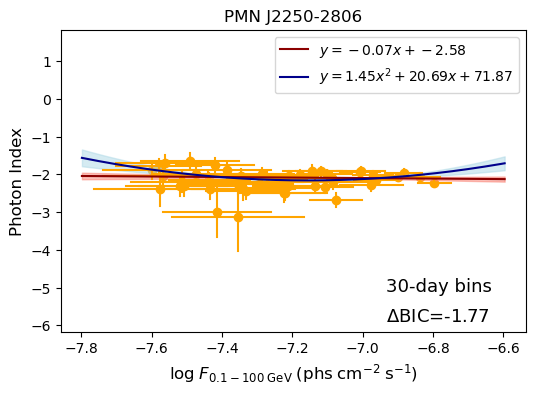}
\includegraphics[width=0.267\linewidth]{3C_454.3.png}
\includegraphics[width=0.267\linewidth]{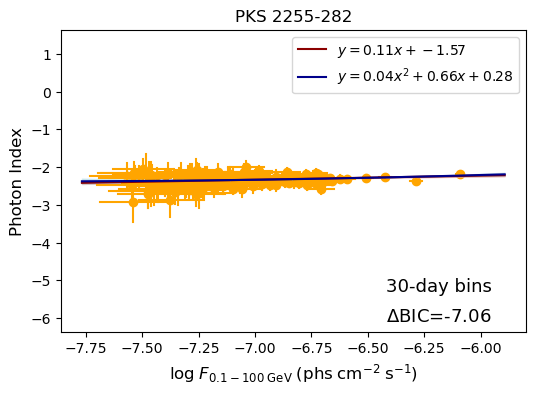}
\includegraphics[width=0.267\linewidth]{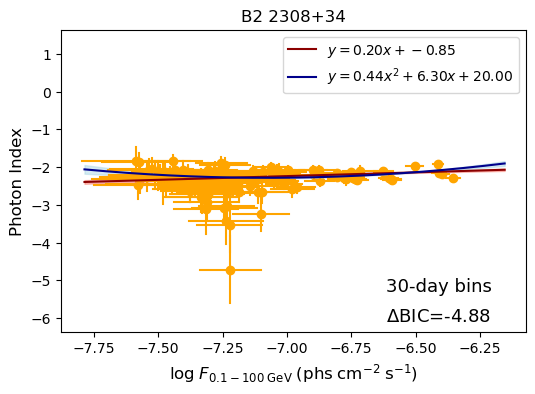}
\includegraphics[width=0.267\linewidth]{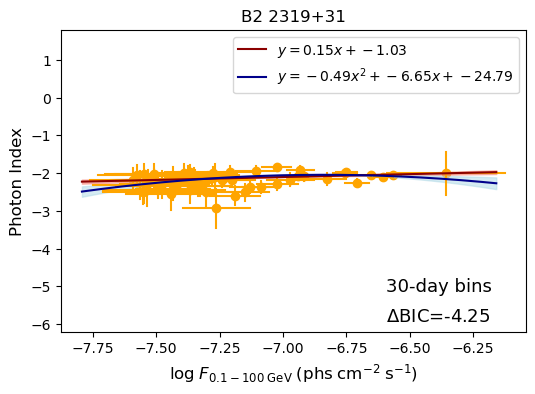}
\includegraphics[width=0.267\linewidth]{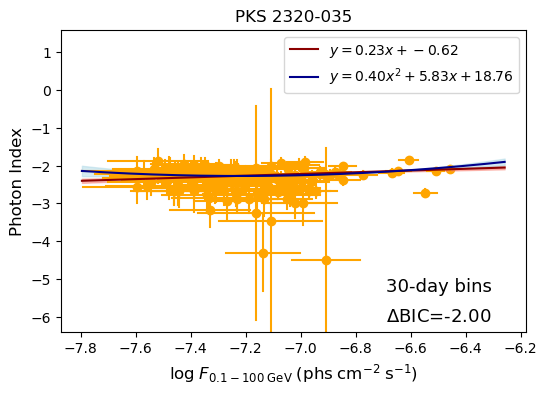}
\includegraphics[width=0.267\linewidth]{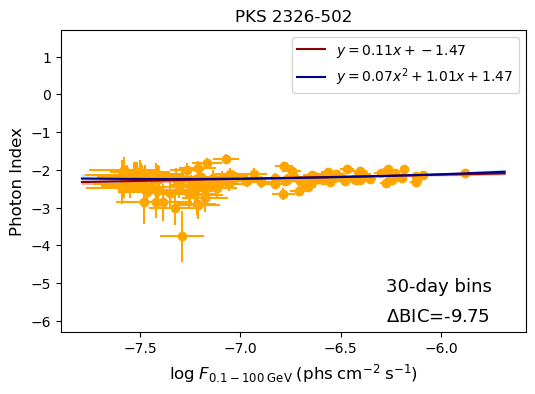}
\includegraphics[width=0.267\linewidth]{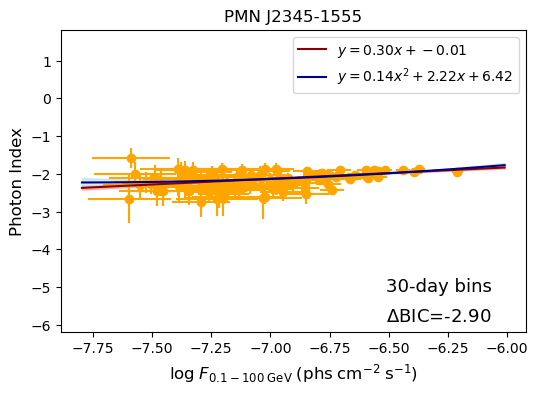}
    \caption{The same as in Figure \ref{fig:indexplot}.} 
    \label{fig:a9}
\end{figure}

Figure \ref{fig:a1}-\ref{fig:a9} shows the 30-day binned photon index plotted against the 30-day binned 0.1-100 GeV flux for all 160 blazars in the sample. 
As the same in Figure \ref{fig:indexplot}, the orange points represent the data with TS values greater than 10. The red line represents the fit of the linear function and the blue line represents the fit of the quadratic function. $\Delta \rm BIC$ is the compared result of the BIC between the two fit functions.

\newpage

\bibliography{sample631}
\bibliographystyle{aasjournal}



\end{document}